\documentclass[fleqn,usenatbib]{mnras}

\usepackage{newtxtext,newtxmath}

\usepackage[T1]{fontenc}

\DeclareRobustCommand{\VAN}[3]{#2}
\let\VANthebibliography\thebibliography
\def\thebibliography{\DeclareRobustCommand{\VAN}[3]{##3}\VANthebibliography}

\usepackage{bm}
\usepackage{graphicx}
\usepackage{multirow,colortbl}
\usepackage{verbatim}
\usepackage{enumitem}
\usepackage{soul}

\usepackage[svgnames]{xcolor}
\newcommand{\msun}{M$_\odot$}

\title[DarkLight: Isolated Dwarfs]{EDGE: Predictable Scatter in the Stellar Mass--Halo Mass Relation of Dwarf Galaxies}

\author[S. Y. Kim et al.]{
Stacy Y. Kim$^{1,2}$\thanks{E-mail: skim11@carnegiescience.edu (SYK)},
Justin I. Read$^{2}$,
Martin P. Rey$^{3}$,
Matthew D. A. Orkney$^{4,5}$,
Sushanta Nigudkar,\newauthor
Andrew Pontzen$^{6}$,
Ethan Taylor$^{2}$,
Oscar Agertz$^{7}$,
Payel Das$^{2}$
\\
$^1$Carnegie Theoretical Astrophysics Center, Carnegie Observatories, 813 Santa Barbara St, Pasadena, CA 91106, USA\\
$^2$Department of Physics, University of Surrey, Guildford, GU2 7XH, UK\\
$^3$Sub-department of Astrophysics, University of Oxford, DWB, Keble Road, Oxford OX1 3RH, UK\\
$^4$Institut de Ciencies del Cosmos (ICCUB), Universitat de Barcelona, Mart\'i i Franqu\`es 1, E-08028 Barcelona, Spain\\
$^5$Institut d'Estudis Espacials de Catalunya (IEEC), E-08034 Barcelona, Spain\\
$^6$Department of Physics and Astronomy, University College London, London WC1E 6BT, UK\\
$^7$Lund Observatory, Division of Astrophysics, Department of Physics, Lund University, Box 43, SE-221 00 Lund, Sweden\\
}

\date{Accepted XXX. Received YYY; in original form ZZZ}

\pubyear{2023}

\begin{document}
\label{firstpage}
\pagerange{\pageref{firstpage}--\pageref{lastpage}}
\maketitle

\begin{abstract}
The stellar-mass--halo-mass (SMHM) relation is central to our understanding of galaxy formation and the nature of dark matter. However, its normalisation, slope, and scatter are highly uncertain at dwarf galaxy scales. In this paper, we present \texttt{DarkLight}, a new semi-empirical dwarf galaxy formation model designed to robustly predict the SMHM relation for the smallest galaxies. \texttt{DarkLight} harnesses a correlation between the mean star formation rate of dwarfs and their peak rotation speed---the $\langle$SFR$\rangle$-$v_{\rm max}$ relation---that we derive from simulations and observations.  Given the sparsity of data for isolated dwarfs with $v_{\rm max} \lesssim 20$\,km/s, we fit the $\langle$SFR$\rangle$-$v_{\rm max}$ relation to observational data for dwarfs above this velocity scale and to the high-resolution EDGE cosmological simulations below. Reionisation quenching is implemented via distinct $\langle$SFR$\rangle$-$v_{\rm max}$ relations before and after reionisation. We find that the SMHM scatter is small at reionisation, $\sim$0.2 dex, but rises to $\sim$0.5 dex ($1\sigma$) at a halo mass of $\sim$10$^9$\,M$_\odot$ as star formation is quenched by reionisation but dark matter halo masses continue to grow. While we do not find a significant break in the slope of the SMHM relation, one can be introduced if reionisation occurs early ($z_{\rm quench} \gtrsim 5$). Finally, we find that dwarfs can be star forming today down to a halo mass of $\sim$2 $\times 10^9$\,M$_\odot$. We predict that the lowest mass star forming dwarf irregulars in the nearby universe are the tip of the iceberg of a much larger population of quiescent isolated dwarfs.
\end{abstract}



\section{Introduction}
\label{sec:intro}
The connection between dwarf galaxies and their dark matter haloes is key to many fundamental questions in astrophysics. This includes the lowest mass halo that can host a galaxy \citep{Jethwa2018}, how feedback regulates star formation in dwarf galaxies \citep[e.g.][]{Agertz2020}, how many dwarfs should be observable in upcoming surveys \citep{2018ApJ...856...69D}, how much dwarfs contribute to reionizing the universe \citep{2017MNRAS.469L..83W}, and to what degree accreted dwarfs can shape the stellar haloes of Milky Way analogs \citep{2022MNRAS.510.4208R}. The galaxy-halo connection is also key for testing dark matter models with galaxy counts.  Indeed, the long-standing `missing satellites problem'---a mismatch between the number of visible dwarf galaxies and the number of dark matter haloes predicted in the standard cold dark matter paradigm ($\Lambda$CDM; \citealt{Klypin99,Moore99})---can be solved with one's choice of the SMHM relation \citep[e.g.][]{Read2017, Jethwa2018, Kim2018, Read2019, Nadler2020}.

However, the relation between galaxies and their dark matter haloes is not well understood at the mass scale of dwarf galaxies, particularly below halo masses $\lesssim 10^{10}$ \msun\ \citep[e.g.][]{Garrison-Kimmel2017, Read2017}.  Many attempts have been made to quantify the simplest form of this relation, in which the stellar mass of a galaxy is a monotonically related to its dark matter halo mass---the stellar mass-halo mass relation (SMHM).  Attempts to match observed galaxies to dark matter haloes via the `abundance matching' approach \citep[e.g.][]{Vale2004} works well for more massive galaxies \citep[e.g.][]{Behroozi2013}.  However, at dwarf scales, such models require the addition of significant stochasticity, which is not well understood \citep[e.g.][]{Garrison-Kimmel2017, Kim2018, Nadler2020}. An alternative approach is to simulate the SMHM from first principles.  However, this comes with its own challenges.  The smallest stellar systems can have stellar masses approaching just $\sim$1000 \msun\ and sizes $\sim$10\,pc \citep{Simon19}, yet must be simulated over cosmologically representative volumes ($\gtrsim50$\,Mpc$^3$) to obtain well-sampled galaxy population statistics. To date, no simulation has been able to capture this dynamic range, leaving a choice between large volume, lower resolution simulations such as Illustris and EAGLE \citep{Genel14,Schaye15}, or more targeted high resolution simulations that `zoom in' on individual galaxies but lack full galaxy population statistics, such as EDGE \citep[e.g.][]{Chan2015, Wang2015, Jeon2017, Munshi2017, Agertz2020, Applebaum2021, Gutcke2022}. It is perhaps not surprising, then, that current simulation results differ by two orders of magnitude below a halo mass of $10^{10}$\,\msun\ (see Figure \ref{fig:halo_occupation}). 
There is also no consensus on the amount of scatter in such relations, with some predicting scatter of an order of magnitude, while others predict far less. Furthermore, most struggle to explain at least some of the observations. Recent high resolution simulations are now able to capture star-forming galaxies at a mass scale of $\sim 5 \times 10^9$\,M$_\odot$ \citep{Jeon2017, Rey2020, Applebaum2021, Gutcke2022}, but none so far have a stellar mass as high as some isolated nearby dwarf irregulars such as Aquarius ($M_* \simeq 6.8 \times 10^5$ \msun) and CVnIdwA \citep[$M_* \simeq 4.1 \times 10^6$ \msun; ][]{Read2017}. It is not clear if this owes to a need to move to yet higher resolution, missing physics, a lack of population statistics, or difficulties in modelling and interpreting the data.

Despite these challenges, recent work has uncovered at least some of the likely sources of scatter in the SMHM relation at low mass. Using a novel ``genetic modification" technique \citep{Roth2016}, \citet{Rey2019} showed, for the faintest dwarfs, the stellar mass is determined by how quickly the halo grows before reionisation, and can produce changes in its final stellar mass by as much as an order of magnitude. The interplay between reionisation and the scatter in dark matter assembly histories is thus a potentially significant contributor to the scatter in the SMHM at its faintest end. At the same time, \citet{Read2019} argued that additional scatter will be present for satellite dwarf galaxies if they quench (i.e. cease forming stars) on infall and/or experience significant mass loss through tides (see also \citealt{Ural15,Tomozeiu16} for similar arguments). They argue that tidally-induced scatter is only significant if stellar mass loss is significant, which is rare\footnote{Evidence for tidal stripping has been reported for a few Milky Way dwarfs \citep[e.g.][]{Ibata01,Pace22,deleo23,Ou24}. Of these, only the Sagittarius dwarf shows evidence for significant stellar mass loss. However, even in this case we can correct for this stellar mass loss using Sagittarius's detected tidal debris \citep{Gibbons17}.}, while the scatter due to quenching can also be removed if abundance matching is performed with a galaxy's mean star formation rate, $\langle$SFR$\rangle$, averaged over the time the dwarf formed stars, rather than its stellar mass, $M_*$.  Using data for nearby satellite dwarfs, they show that $\langle$SFR$\rangle$ correlates better with halo mass, M$_{200}$, than does M$_*$.

In this paper, we build on the above findings to present a new, fast, empirical galaxy-halo model, \texttt{DarkLight}.  We improve on the $\langle$SFR$\rangle$-$M_{200}$ correlation presented in \citet{Read2019},   
showing that for very low mass dwarfs, $\langle$SFR$\rangle$ correlates more strongly with the peak rotation speed of a dwarf, $v_{\rm max}$, than $M_{200}$. We then harness this new $\langle$SFR$\rangle$-$v_{\rm max}$ relation to turn individual dark matter halo assembly histories, $v_{\rm max}(t)$, into star formation histories (SFHs). Integrating the SFHs then returns the stellar mass of the dwarf, $M_*(t)$. We show that, once calibrated, this simple mapping is able to reproduce the stellar mass growth history of each of our simulated EDGE dwarfs over time. As we apply the mapping to all progenitors of the dwarf and track any accreted material, we are also able to reproduce any sudden jumps in stellar mass due to dry mergers. 

\texttt{DarkLight} can be fully calibrated on measurements of $\langle$SFR$\rangle$ and $v_{\rm max}$ from nearby dwarf galaxies. This means that, with sufficiently good calibration data in hand, \texttt{DarkLight} need not rely on uncertain galaxy formation simulations, opening the door to robust cosmological constraints from dwarf galaxy number counts. At present, however, a statistically significant sample of data is only available for dwarf galaxies with $v_{\rm max} \gtrsim 20$ km/s. As such, here we calibrate \texttt{DarkLight} on observations above this velocity scale, and on the EDGE simulations below.

This paper is organised as follows. In Section \ref{sec:darklight}, we present the details of \texttt{DarkLight}. In Section \ref{sec:validation}, we validate \texttt{DarkLight} by showing that it accurately reproduces the stellar mass evolution of individual EDGE dwarfs.  Once calibrated, \texttt{DarkLight} only requires the growth history of a dark matter halo, and thus can be run on computationally inexpensive dark-matter-only (DMO) simulations of large cosmological volumes to predict the properties of large, statistical samples of dwarfs, down to the very smallest stellar systems. To illustrate this, in Section \ref{sec:smhm}, we run \texttt{DarkLight} on a large, cosmological void volume (50\,Mpc$^3$), to derive the SMHM relation {\it as if} we had run the entire volume with the resolution and baryonic physics model of EDGE.  We study the origins of SMHM scatter and discuss the role of reionization in shaping the SMHM and its scatter and its sensitivity to changes in \texttt{DarkLight}'s input parameters.  In Section \ref{sec:starform}, we study whether fully sampling over assembly histories in a large cosmological volume is sufficient to explain puzzling low mass star-forming galaxies in the nearby Universe such as Aquarius and CVnIdwA.  We discuss our results, compare \texttt{DarkLight} against other galaxy formation models and hydrodynamic simulations, and outline its limitations in Section \ref{sec:discussion}. Finally, in Section \ref{sec:conclusions}, we present our conclusions.


\section{DarkLight}
\label{sec:darklight}

Dwarf galaxies represent the extreme end of galaxy and star formation.  As such, they remain a challenge to model numerically and are highly sensitive to the assumed subgrid physics \citep[e.g.][]{Munshi2019, Agertz2020}.  Galaxy formation models calibrated to reproduce massive galaxies often fail when run in the dwarf regime without modification  (e.g. discussion in Sec. \ref{sec:discussion}), indicating that different processes become important at low-mass scales.  Here, we present a new model, \texttt{DarkLight}, that is tailored to accurately predict dwarf galaxy properties.  \texttt{DarkLight} is based on two recent findings regarding the formation of dwarfs: (i) that, for reionisation-quenched dwarfs, dark matter assembly histories are particularly important in determining their stellar content \citep{Rey2019}; and (ii) that their star formation rates correlate with halo properties \citep{Read2019}.  In Section \ref{sec:darklight:sfr-vmax}, we discuss the latter and present a new $\langle{\rm SFR}\rangle$-$v_{\rm max}$ relation based on observed and simulated dwarf galaxies upon which \texttt{DarkLight} is based. In Section \ref{sec:darklight:mapping}, we discuss how we apply this relation to the assembly histories of dwarf haloes to predict their stellar content over time. \texttt{DarkLight} is open-source and available to download at \href{https://github.com/stacykim/darklight}{https://github.com/stacykim/darklight}.


\subsection{The $\langle$SFR$\rangle$-$v_{\rm max}$ relation at dwarf scales}
\label{sec:darklight:sfr-vmax}

\begin{figure}
    \centering
    \includegraphics[width=0.5\textwidth]{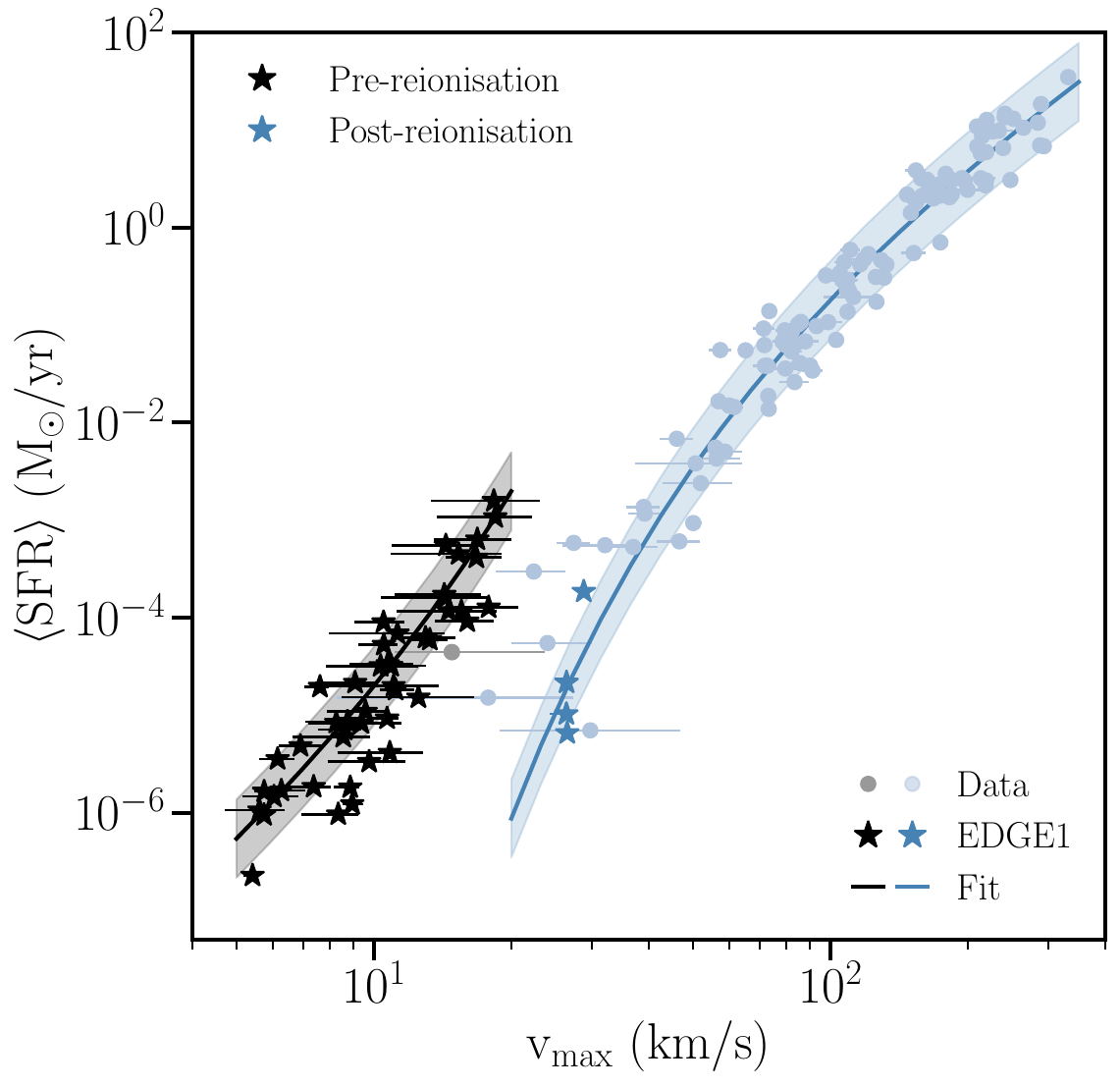}
    \caption{Relation between the average star formation rate and $v_{\rm max}$. We find two relations: one pre-reionisation, shown in black, and another, with suppressed star formation rates, post-reionisation, shown in blue. The circles are derived from observational data (see Sec. \ref{sec:darklight:sfr-vmax} for details), while the stars are from the high-resolution cosmological EDGE suite of simulations. The dwarf galaxy represented by the gray circle just above $v_{\rm max} \sim 10$\,km/s is Eridanus II.  In effect, the post-reionisation relation is calibrated by observational data, while the pre-reionisation relation is calibrated on the EDGE simulations.}
    \label{fig:sfr-vmax}
\end{figure}

\citet{Read2019} showed that mean star formation rate, $\langle{\rm SFR}\rangle$, of the Milky Way's classical dwarf galaxies, and nearby dwarf irregulars, correlates tightly with halo mass, $M_{200}$. We find that there is considerably less scatter when relating $\langle$SFR$\rangle$s to the maximum of the rotation curve, $v_{\rm max}$ (often used as a proxy for mass; e.g. \citealt{Boylankolchin12}) rather than halo mass, $M_{\rm 200}$ (see Appendix \ref{appendix:M200}).  This is due to two main effects. Firstly, $v_{\rm max}$ is better constrained in observations and less susceptible to mergers and environmental effects. Secondly, the star formation rate better correlates with the inner mass distribution (since this is where the gas and stars reside), which is better constrained by $v_{\rm max}$ than $M_{200}$. For these reasons, in this paper we use the correlation between $\langle{\rm SFR}\rangle$ and $v_{\rm max}$, rather than the weaker correlation between $\langle{\rm SFR}\rangle$ and $M_{200}$.

As reionisation suppresses or even halts star formation in dwarf galaxies, we construct a separate $\langle{\rm SFR}\rangle$-$v_{\rm max}$ relation before and after reionisation quenching. To construct the post-reionisation relation, we largely rely on data from observed nearby isolated galaxies. For galaxies with $v_{\rm max} \gtrsim $ 20 km/s, we use the observational sample of (non-rogue) dwarf irregulars in \citet{Read2017}, which have stellar masses and $v_{\rm max}$ derived from photometric light profiles and HI rotation curves. A few of these galaxies have detailed star formation histories derived from deep color-magnitude diagrams. These include the dwarfs WLM \citep{2019MNRAS.tmp.2515A}, Aquarius \citep{Cole2014}, Leo T \citep{2012ApJ...748...88W}, and IC 1613\footnote{Note that IC1613 is actually classed as a rogue in \cite{Read2017}. However, we can use it here since, combining is stellar and HI gas kinematics, its inclination is determined robustly, allowing for a trustworthy inference of $v_{\rm max}$ \citep{Oman16, Collins22}.} \citep{2014ApJ...786...44S}. For these four galaxies, we calculated their $\langle$SFR$\rangle$s averaged over the last 1\,Gyr. For the remainder of the dwarfs, that do not have detailed star formation histories, we calculated their mean lifetime star formation rates $\langle$SFR$\rangle = M_* / t_{\rm H}$, where $M_*$ is the stellar mass and $t_{\rm H}$ the Hubble time, based on stellar masses derived by \citet{2012AJ....143...47Z} from SED fitting (c.f. the similar analysis adopted in \citealt{Read2017} and \citealt{Read2019b}).

We supplement these dwarfs with the SPARC dataset of disk galaxies. As detailed star formation histories are not available for these galaxies, we again compute their mean lifetime star formation rates based on stellar masses derived from their 3.6 $\mu$m luminosities published in \citet{2016AJ....152..157L} and stellar mass-to-light ratios published in \citet{2019A&A...626A..56P}. For $v_{\rm max}$, we use $v_{\rm flat}$ calculated by \citet{2016AJ....152..157L} based on HI/H$\alpha$ rotation curves. 

Observational data on dwarfs with $v_{\rm max} \lesssim$ 20 km/s is sparse. While there are many observed dwarfs below this threshold, most are satellite galaxies, for which quenching and tidal stripping may have significantly affected their star formation histories. Restricting to isolated dwarf galaxies leaves us with one good candidate, Eridanus II \citep{Li17}. Eridanus II has a very old stellar population that appears to have been quenched at about $t_{\rm quench}$ = 2\,Gyr \citep{Simon21}.  We computed its $\langle$SFR$\rangle$ = M$_*^{\rm Eri II}$ / $t_{\rm quench}$ = $4.44 \times 10^{-5}$ \msun/yr.  As its star formation history is consistent with being a reionisation relic, we used it in our fits for the pre-reionisation relation.

To supplement these data, we compiled $\langle$SFR$\rangle$ and $v_{\rm max}$ from the Engineering Dwarfs at Galaxy formation's Edge (EDGE) simulation suite of low-mass dwarfs with halo masses $M_{200} = 1-4 \times 10^9$~\msun\ \citep{Agertz2020, Rey2019, Orkney2021}. These fully cosmological zoom-in simulations are run with the adaptive mesh refinement code \textsc{RAMSES} \citep{Teyssier2002} and adopt the cosmological parameters $\Omega_M = 0.309$, $\Omega_\Lambda = 0.691$, $\Omega_b = 0.045$, and $H_0 = 67.77$ km/s/Mpc \citep{Planck2014}. They have a maximum spatial resolution of 3\,pc and self-consistently resolve the momentum and energy deposition from individual supernovae explosions. Dark matter is represented by particles with a mass resolution of 960\,M$_\odot$. Reionisation heating is modeled via an updated version of \citet{Haardt1996} with a high-redshift cutoff implemented in the publicly available version of \textsc{RAMSES}.  Reionisation begins at $z \sim 6.5$, but star formation fully quenches at $z_{\rm quench} = 4$. A more complete description of the subgrid physics and the EDGE simulations is given in \citet{Agertz2020} and \citet{Rey2020}.  Halos are identified using the HOP halo finder \citep{Eisenstein1998}, and merger trees constructed by matching halos between snapshots using \texttt{pynbody} \citep{Pontzen2013} and \texttt{tangos} \citep{Pontzen2018}.  Halo masses $M_{200}$ are defined as the mass within a radius $r_{200}$ that encloses an average density that is 200 times the critical density of the universe.

For EDGE, we compute pre-reionisation values for all isolated galaxies that exist in our high-resolution volume at $z_{\rm quench}$. We average the (in-situ) star formation rate and $v_{\rm max}$ from the birth times of each halo to $z_{\rm quench}$. The $v_{\rm max}$ is computed at each time step from rotation curves based on all particles in the simulation.  For those haloes that form stars again \citep[i.e. ``rejuvenate",][]{Wright2019, Rey2020} after reionisation, we compute the average star formation rate and $v_{\rm max}$ from the time they rejuvenate to the present day, and use these to fit the post-reionisation relation.  There are more pre-reionisation than post-reionisation data points from EDGE due to the fact that there are many isolated galaxies before reionisation (many of which eventually merge with the main progenitor by $z=0$), and because few of the EDGE dwarfs are rejuvenators.

Fits to the resultant dataset are shown in Fig. \ref{fig:sfr-vmax}. Observational data are represented by circles, while simulation data are represented by stars. The pre-reionisation relation is shown in grey, while the post-reionisation relation is shown in blue. Where the EDGE and the observational data overlap, they agree well. The best-fit pre-reionisation relation is given by:
\begin{equation}
    \langle {\rm SFR} \rangle = 2 \times 10^{-7} \left( \frac{v_{\rm max}}{5} \right)^{3.75} e^{v_{\rm max}/5} \qquad {\rm M}_\odot ~ {\rm yr}^{-1}
\end{equation}
with a 1$\sigma$ symmetric scatter of 0.4\,dex. Notice that the relation flattens slightly towards low $v_{\rm max}$. The best-fit post-reionisation relation is given by:
\begin{equation}
    \langle {\rm SFR} \rangle = 7.06 ~ \left( \frac{v_{\rm max}}{182.4} \right)^{3.07} e^{-182.4/v_{\rm max}} \qquad {\rm M}_\odot ~ {\rm yr}^{-1}
\end{equation}
with a 1$\sigma$ symmetric scatter of 0.3\,dex. These two relations are shown in Fig. \ref{fig:sfr-vmax} as black and blue lines, respectively, with the $1\sigma$ scatter marked by bands.


\subsection{Constructing dwarf galaxies}
\label{sec:darklight:mapping}

We can now apply the $\langle$SFR$\rangle$-$v_{\rm max}$ relation from Section \ref{sec:darklight:sfr-vmax} to generate star formation histories, SFR$(t)$, and stellar masses, $M_*(t)$, from the $v_{\rm max}(t)$ trajectory of a halo drawn from a dark matter only simulation. We correct the $v_{\rm max}(t)$ trajectory to subtract the baryonic contribution (much of which is evaporated away for the lowest-mass dwarfs during reionisation) by multiplying by $\sqrt{1-f_b}$, where $f_b = \Omega_b/\Omega_m$ = 0.17 is the baryon fraction of the universe. In Appendix \ref{appendix:vmaxDMO}, we show that this provides a better match to the $v_{\rm max}$ the halo would have had if it had been run in a hydrodynamic simulation. We take the average $v_{\rm max}$ before reionisation and use the pre-reionisation $\langle$SFR$\rangle$-$v_{\rm max}$ relation (the relation in black/gray in Fig. \ref{fig:sfr-vmax}) to derive a single, average pre-reionisation SFR  
before reionization quenching, $z < z_{\rm quench}$. After $z > z_{\rm quench}$, haloes do not form stars unless/whenever they pass above a threshold $v_{\rm max}^{\rm post}$, upon which we assign a single, average SFR based on the halo's $v_{\rm max}$ at $z$=0 from the post-reionisation relation (the relation in blue in Fig. \ref{fig:sfr-vmax}). For each timestep in which star formation occurs after reionisation, we add scatter from the appropriate $\langle$SFR$\rangle$-$v_{\rm max}$ relation.

The above procedure produces a stellar mass trajectory for stars formed in-situ, $M_*(t)$. In addition, we calculate the contribution from {\it accreted} haloes by running {{\tt DarkLight}} on all haloes that merge onto the main halo. Optionally, one can impose a halo occupation function to capture the suppression of galaxy formation in low-mass halos, which have limited ability to accrete and cool star-forming gas, particularly after reionisation.  In this case, whether each accreted halo contains a galaxy or not is probabilistically determined from the halo occupation function.  By default, \texttt{DarkLight} does not adopt a halo occupation function.

For any given halo, \texttt{DarkLight} can generate multiple realisations of the halo's stellar mass growth history, sampling the scatter in the $\langle$SFR$\rangle$-$v_{\rm max}$ relation and yielding confidence intervals on $M_*(t)$.


\begin{figure*}
\begin{center}
\includegraphics[width=\textwidth]{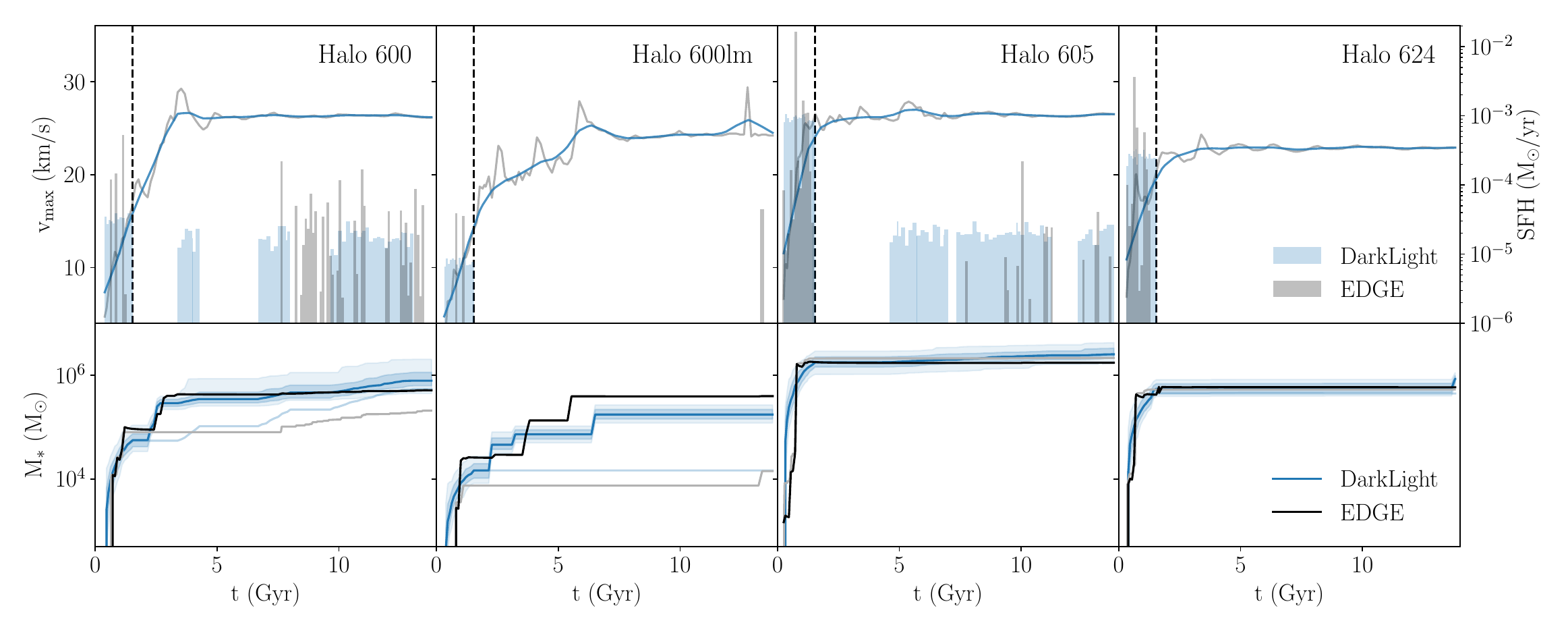}
\includegraphics[width=\textwidth]{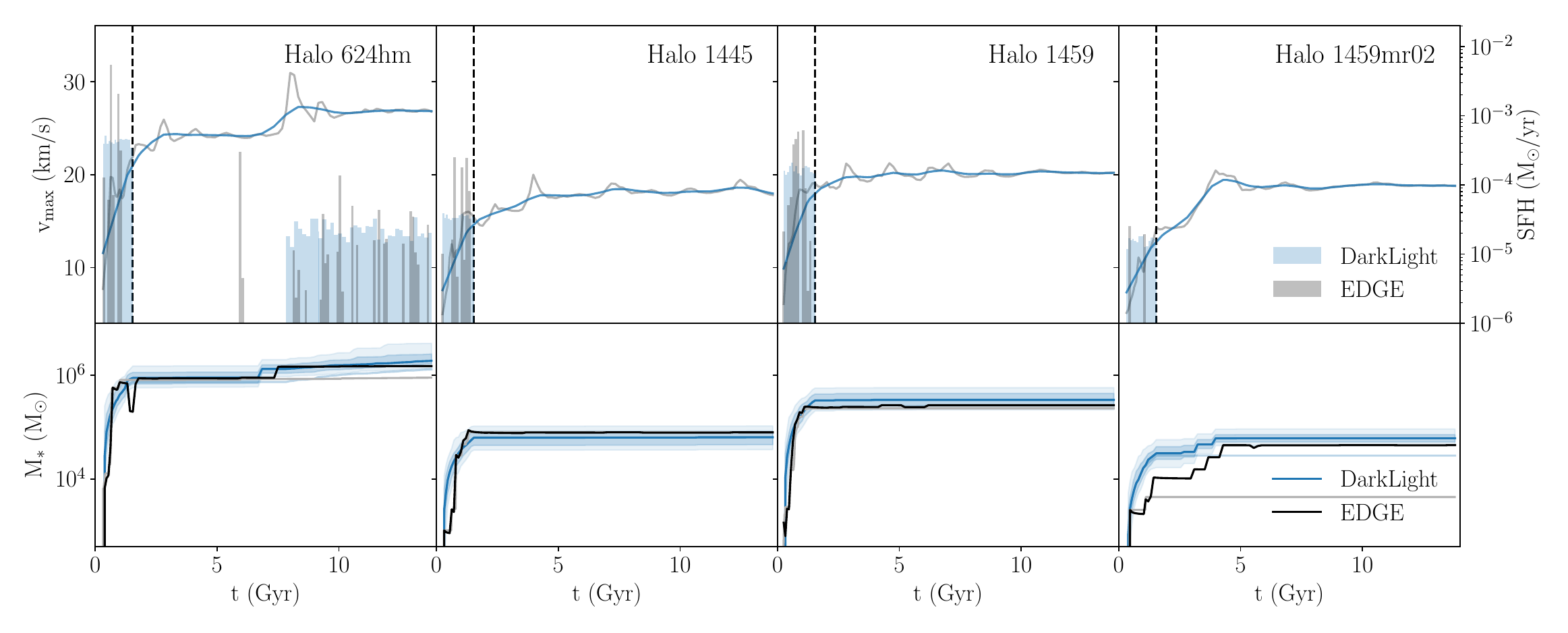}
\caption{\texttt{DarkLight}'s prediction for the star formation histories and stellar mass trajectories, $M_*(t)$, for several EDGE dwarfs. In each row, the lines in the top panels denote the $v_{\rm max}$ evolution of different haloes.  The $v_{\rm max}$ trajectories from the EDGE simulations is shown in gray.  These trajectories are smoothed (blue lines) before being fed into \texttt{DarkLight}. Dashed black vertical lines denote $z_{\rm quench}$ = 4, when reionisation shuts off star formation in the lowest mass EDGE dwarfs. Star formation histories from EDGE and \texttt{DarkLight} are overplotted as gray and blue histograms, respectively. The bottom panels of each row show the cumulative stellar mass growth of several EDGE dwarf galaxies predicted by \texttt{DarkLight} (blue) compared to EDGE simulations of each (black).  The black and dark blue lines denote the total stellar mass growth (in-situ and accreted), while the gray and light blue lines denote in-situ stellar mass growth (note that the in-situ mass trajectories do not include stellar mass loss, and thus can be higher than the total mass).  The dark and light blue bands on the \texttt{DarkLight} stellar mass trajectories show the 1- and 2-$\sigma$ range based on 100 \texttt{DarkLight} realisations of each halo's SFR, accounting for the scatter in the SFR-$v_{\rm max}$ relation. Note that the stellar masses predicted by \texttt{DarkLight} are accurate to within 2$\sigma$, and within a factor of two of the hydrodynamic simulations for all cases. The worst recovery is for late-forming haloes (Halo 600lm and Halo 1459mr02) that are built up by many small mergers ($M_{200} \lesssim 10^8$ \msun) for which the $v_{\rm max}$ are difficult to measure.}
\label{fig:validation}
\end{center}
\end{figure*}

\section{Validation: reproducing hydrodynamic simulations}
\label{sec:validation}

In this section, we show how \texttt{DarkLight} reproduces the stellar mass growth history, $M_*(t)$, of the fully hydrodynamic, high-resolution dwarf galaxies in the EDGE suite. To do this, we first extracted the $v_{\rm max}(t)$ trajectories for each of the dwarfs from DMO zoom simulations. We then smoothed the trajectories with a 1\,Gyr Gaussian filter to remove small fluctuations in $v_{\rm max}(t)$ that occur due to mergers/interactions and calculated SFR$(t)$ based on the smoothed trajectories using \texttt{DarkLight}. 

The results are shown in Fig. \ref{fig:validation}. We find that $v_{\rm max}$ = 26.3 km/s and $z_{\rm quench}$ = 4 reproduce the EDGE dwarfs well. The upper panels of each row show the original $v_{\rm max}(t)$ trajectories from EDGE with a gray line.  These are smoothed (blue line) before being input to \texttt{DarkLight}. The star formation histories from EDGE and \texttt{DarkLight} are overplotted as gray and blue histograms, respectively. The bottom panels show the evolution of the total stellar mass in EDGE versus \texttt{DarkLight} in black and dark blue lines, respectively. The stellar mass formed in-situ in EDGE versus \texttt{DarkLight} is shown in gray and light blue lines, respectively. The bands denote the 1 and 2$\sigma$ scatter in the range of predicted stellar masses based on 100 realizations of each halo. For most haloes, the median $M_*(t)$ produced by \texttt{DarkLight} matches the stellar mass trajectories in EDGE within 2$\sigma$. This includes an excellent recovery of reioinisation-quenched dwarfs like Halo 1445 and Halo 1449, and `rejuvenators' that reignite their star formation sometime after reionisation like Halo 600 and Halo 605 \citep[see also][for a discussion of rejuvenation]{Rey2020}.

However, late-forming haloes that are built up by many small mergers (typically $\sim$2$\sigma$ outliers) are less well reproduced (e.g. Haloes 600lm and 1459mr02).  This is due to the difficulty of reproducing the stellar masses of the very low mass haloes ($M_{200} \lesssim 10^{8}$ \msun) they merge with, which is in turn caused by difficulties in measuring $v_{\rm max}$ for such haloes. However, even in these scenarios, the final stellar masses are reproduced to within a factor of a few. We discuss how we can improve upon this in future work in Sec \ref{sec:disc:improvements}.


\section{The Stellar-Mass--Halo-Mass Relation}
\label{sec:smhm}

\begin{figure}
    \centering
    \includegraphics[width=0.5\textwidth]{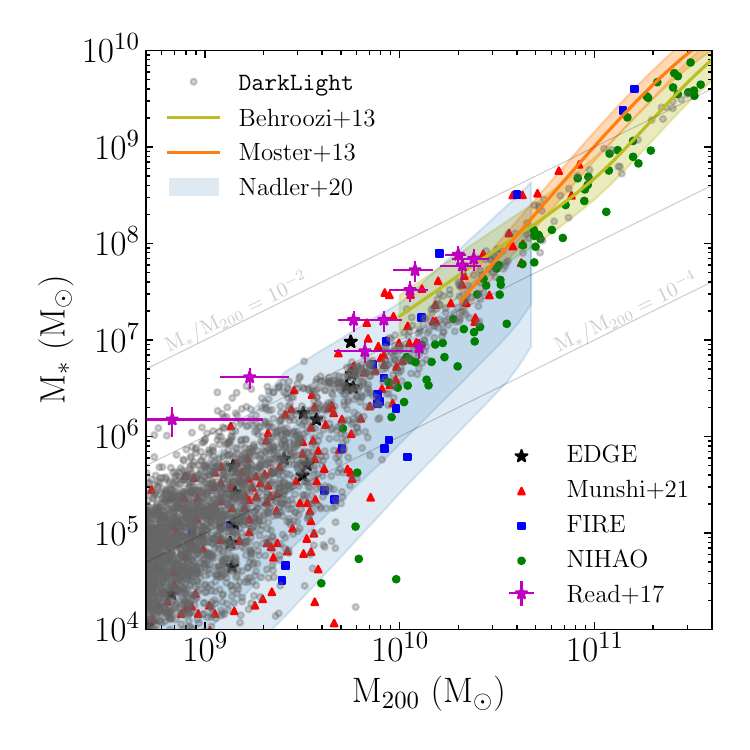}
    \caption{The SMHM relation predicted by \texttt{DarkLight} for haloes in EDGE's parent void volume. \texttt{DarkLight} predictions are plotted in gray. Overplotted are dwarf galaxies from a variety of simulation studies, including EDGE \citep[black stars;][]{Agertz2020, Rey2020, Orkney2021}, the DC Justice League and Marvelous Dwarf suites \citep[red triangles;][]{Munshi2021}, FIRE \citep[blue squares;][]{Chan2015, Wheeler2015, Wetzel2016, Fitts2017}, and NIHAO \citep[green circles;][]{Wang2015}.  Observed gas-rich dwarfs from \citet{Read2017}, which were used as part of the calibration data for \texttt{DarkLight}, are shown in magenta.  These observed dwarfs are likely a biased subsample of low-mass dwarfs, representing the brightest members.  At $M_{200} \gtrsim 10^{10}$ \msun, \texttt{DarkLight} is in good agreement with abundance matching relations such as \citet{Behroozi2013} and \citet{Moster2013}, which are plotted with yellow and orange lines, respectively.}
    \label{fig:smhm}
\end{figure}

In this section, we present a first application of \texttt{DarkLight} by applying it to the dark-matter-only (DMO) simulation of the void volume from which the EDGE dwarf galaxies were selected. This volume was selected from a DMO cosmological simulation with a box size of 50\,Mpc and a resolution of 512$^3$ particles. We ran \texttt{DarkLight} on all haloes that did not have a merger history indicative of a close encounter with a more massive halo (i.e. that showed no unusual jumps or discontinuities in their mass growth history). In this way, we populate the entire void {\it as if} we had run EDGE on every galaxy in the volume. 

In Fig. \ref{fig:smhm}, we show \texttt{DarkLight}'s prediction for the SMHM relation of this void volume as grey points, plotting the median stellar mass obtained from integrating 100 realizations of the SFH for each halo. Overplotted are results from recent observational, simulation, and abundance matching studies. Observational data from isolated gas-rich dwarf irregulars, with halo masses inferred from their HI rotation curves from \citet{Read2017} are plotted with magenta stars. The black stars show galaxies from the EDGE simulations presented in Fig. \ref{fig:validation}---families of haloes created by genetically modifying a single halo to form earlier or later appear as vertical lines of points. Those from NIHAO \citep{Wang2015}, FIRE \citep{Chan2015, Wheeler2015, Wetzel2016, Fitts2017}, and the Marvel-ous and DC Justice League \citep{Munshi2021} simulation suites are shown as green circles, blue squares, and red triangles, respectively.  Abundance-matching relations from \citet{Behroozi2013} and \citet{Moster2013} are shown by the yellow and orange lines, respectively.

At halo masses $M_{200} \gtrsim 10^{10}$ \msun, the differences between all of the studies, including \texttt{DarkLight}, are small. However, at halo masses $M_{200} \lesssim 10^{10}$ \msun, the differences between the studies grow.  Notably, the simulations predict that galaxies have significantly less stellar mass for a fixed halo mass than the observed galaxies (magenta), with differences as large as two orders of magnitude at the lowest masses. Furthermore, the differences between simulation suites are also large, with disagreement in not just the normalisation but also in the slope and scatter of the SMHM relation \citep[as also observed by][]{Garrison-Kimmel2017}.

In contrast, \texttt{DarkLight} straddles the space between simulations and observations. We discuss possible reasons for the discrepancies between \texttt{DarkLight} and the different simulations in Section \ref{sec:smhm:sims}. \texttt{DarkLight} suggests that the lowest mass HI-rich dwarfs (magenta) are outliers in stellar mass for their halo mass---the tip of the iceberg of a large population of quenched dwarf galaxies with lower stellar masses that remain to be discovered. This implies that abundance matching relations based on this sample of observed dwarfs likely suffer from incompleteness and over- (under-)estimate stellar (halo) masses, as was found for \citet{Behroozi2013}'s relation for higher-mass dwarf galaxies \citep{Garrison-Kimmel2014}.

Some studies have found that the SMHM relation appears to have a ``knee" below which the relation steepens in slope. The location of the knee varies from study to study, ranging from M$_{200}$ = $10^{10}$ \msun~\citep[][for their ``all haloes" SMHM relation]{Munshi2021} to $10^{11}$ \msun~\citep{Nadler2020, Danieli2023}. We find that \texttt{DarkLight} does not predict a clear knee given our fiducial model parameters. However, we show in Section \ref{sec:smhm:sensitivity} that we can generate a stronger knee by shifting the redshift at which dwarf galaxies quench earlier. One can also be generated by decreasing pre-reionisation and/or increasing post-reionisation star formation rates to create a break at this mass scale. This occurs around where the population transitions from star-forming dwarfs (including rejuvenators) to quenched reionisation relics (see Sec. \ref{sec:starform}). Thus if such a knee exists, the location of the knee, as well as the slopes above and below it, could point towards the timing of reionisation quenching, the mass scale at which dwarf galaxies rejuvenate, and differences in star formation rates before and after reionisation.

Finally, \texttt{DarkLight} naturally produces an increase in scatter towards smaller halo masses, as predicted by a number of hydrodynamic simulations \citep[e.g.][]{Sawala2016, Munshi2017}, galaxy-halo models \citep{Manwadkar2022, OLeary2023, Ahvazi2024}, and explored in some abundance matching studies \citep{Garrison-Kimmel2017} and fits to observed nearby dwarf galaxies \citep{Nadler2020, Danieli2023}. The increase in scatter is often attributed to the rise in stochasticity of star formation at the low-mass end of galaxy formation. In contrast, \texttt{DarkLight} predicts that this scatter owes almost entirely to the combination of reionisation quenching and the scatter in the assembly histories of low mass dwarfs and is, therefore, deterministically predictable. We discuss this further in Section \ref{sec:smhm:scatterorigin}. The median relation predicted by \texttt{DarkLight} is given by:
\begin{equation}
\log M_* = 1.76~\log M_{\rm 200} -10.7
\end{equation}
with a increasing scatter described by:
\begin{equation}
\sigma = -0.227~\log M_{\rm 200} + 2.59
\end{equation}
At a halo mass of 10$^9$ M$_\odot$, we measure a 1$\sigma$ symmetric scatter in stellar mass of 0.49\,dex. Similarly, \citet{Nadler2020} predict a similar though slightly larger scatter of $\sim$0.56 dex (taking the average from the posterior) based on the Milky Way satellites.  \citet{Munshi2021} predict a larger scatter of 0.73 dex, which could be due to their halo occupation fraction, which predict haloes with $M_{200} \gtrsim 10^9$ \msun\, are occupied (see Sec. \ref{sec:smhm:sensitivity}).  These are significantly smaller than the 1.4\,dex inferred by \citet{Danieli2023} when fitting an increasing scatter model to nearby dwarf satellite galaxies observed in the ELVES survey (although if they assume a constant scatter, they infer a very small scatter of 0.06 dex).  This may be due to the additional quenching that dwarfs in their sample may experience as satellite galaxies, and/or the wide range of hosts their satellites are drawn from.


\subsection{The origins and predictability of increasing scatter}
\label{sec:smhm:scatterorigin}

\begin{figure}
    \centering
    \includegraphics[width=0.5\textwidth]{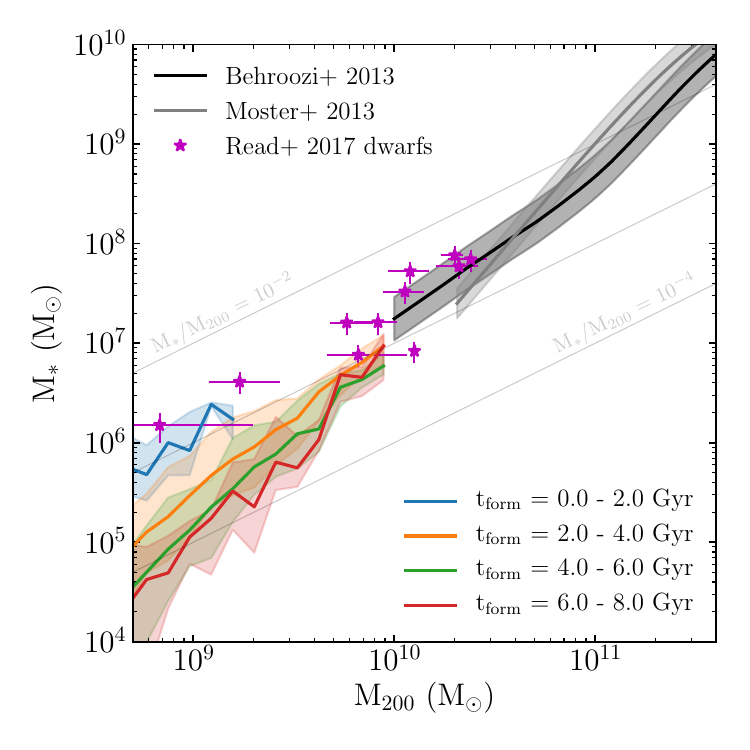}
    \caption{The SMHM relation predicted by \texttt{DarkLight}, broken down by halo formation time.  Haloes have systematically higher stellar masses if they formed earlier for a fixed halo mass. The colored lines denote the median stellar masses for haloes with formation times $t_{\rm form}$ spanning from 0-8\,Gyr. Formation times are defined as the time at which a halo achieves half of its present day mass. The shaded bands represent the 1$\sigma$ scatter in $M_*$ for the haloes in each $t_{\rm form}$ bin.  There are no haloes with $t_{\rm form} < 1.5$\,Gyr above $\sim 10^9$ \msun, causing a truncation in this band.}
    \label{fig:smhm-tform}
\end{figure}

\begin{figure}
    \centering
    \includegraphics[width=0.5\textwidth]{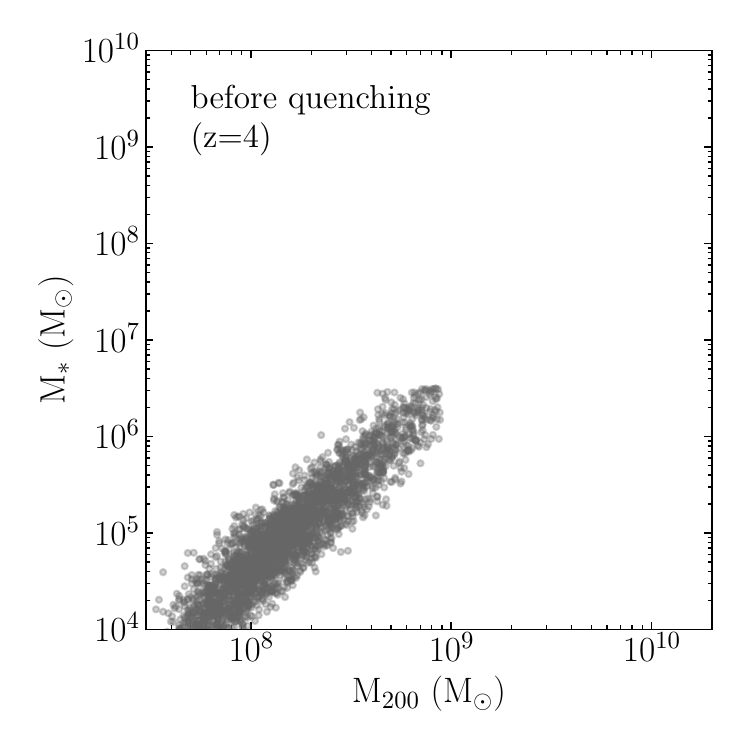}
    \caption{The SMHM at reionisation quenching ($z$=4) predicted by \texttt{DarkLight} for low-mass halos. Unlike \texttt{DarkLight}'s SMHM at $z=0$, this does not exhibit an increase in scatter at low halo masses. The scatter at $<10^9$ M$_\odot$ is roughly constant with a 1$\sigma$ symmetric scatter of 0.23\,dex. This demonstrates that the increasing scatter seen at $z=0$ in low-mass haloes is due to post-reionisation evolution, namely that low mass dark matter haloes can grow in mass, while their stellar mass---quenched by reionisation---cannot (at least through in-situ star formation). The gray band denotes the numerical resolution limit for our simulations.}
    \label{fig:smhm-reionization}
\end{figure}

The increase in scatter at the low-mass end is often attributed to an increase in the stochasticity of star formation in low-mass haloes. While the stochasticity of star formation does play a role, we find that, in the regime probed here, $M_{200} \gtrsim 10^9$ M$_\odot$, the majority of the SMHM scatter can instead be explained by the dark matter growth history of a low-mass halo---and is thus largely predictable.

Two distinct processes produce this increase in scatter. Firstly, the differences in the growth histories of dark matter haloes cause substantial scatter in the SMHM relation at the low mass end, as demonstrated for one EDGE halo that was ``genetically modified" to grow faster or slower \citep{Rey2019}. We find that this holds true across the population of galaxies generated by \texttt{DarkLight} (we note that running on void halos may bias our sample towards later formation times). In Fig. \ref{fig:smhm-tform}, we show the median stellar masses for haloes with different formation times, which we define to be the time when a halo reaches half of its present-day mass. The difference between the medians of the lowest and highest formation time bins is about an order of magnitude at $10^9$ \msun, in agreement with \citet{Rey2019}, and decreases at higher masses. Below M$_{200} \lesssim 10^{10}$ \msun, we find that for a fixed halo mass, the earliest forming haloes tend to host the most massive galaxies, while the latest forming haloes host the least massive galaxies. This implies that early formers tend to sit in the upper envelope of the SMHM relation, while later formers sit in the lower envelope. For many of these galaxies, the vast majority of their stellar mass formed before reionisation. Their haloes assembled a greater fraction of their final mass by reionisation (and thus have a higher pre-reionisation $v_{\rm max}$) and thus were able to form stars at a higher rate before being quenched.

\begin{figure}
    \centering
    \includegraphics[width=0.5\textwidth,trim=0cm 0.65cm 0cm 0cm,clip=true]{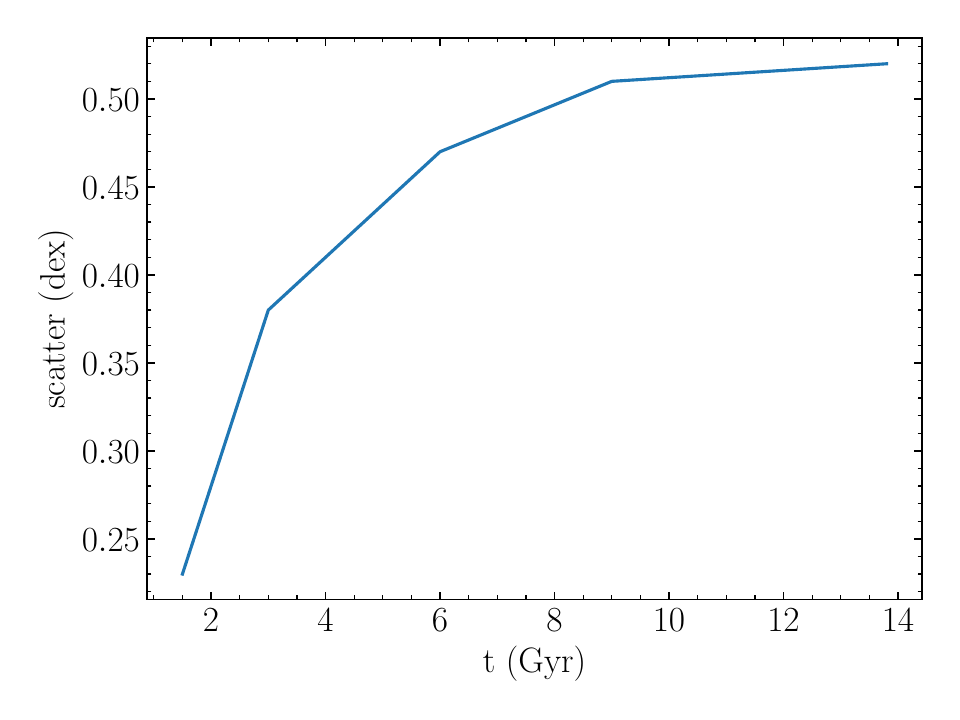}
    \caption{Growth in the SMHM scatter for progenitors of $M_{200}(z=0) = 10^9$ \msun\,haloes over time.  At reionisation quenching, which occurs at $\sim$1.5 Gyr in EDGE and \texttt{DarkLight}, there is a constant 1$\sigma$ scatter of about 0.23 dex, which grows to about 0.5 dex by the present day.}
    \label{fig:scatter-over-time}
\end{figure}

Secondly, after reionisation, low-mass haloes halt star formation until they are sufficiently massive (as captured by our threshold $v_{\rm max}^{\rm post} = 26.3$ km/s in \texttt{DarkLight}). Thus, while before reionisation, the dark matter and stellar masses could grow roughly in step with one another, reionisation decouples the two. Stellar masses can remain largely stagnant after reionisation, while their dark matter haloes can continue to grow. As the growth in dark matter mass can vary for different haloes, this produces an increase in SMHM scatter over time.  In Fig. \ref{fig:smhm-reionization}, we show the SMHM predicted by DarkLight when reionisation quenching occurs ($z \sim 4$ in EDGE). Unlike the SMHM at $z=0$, the scatter is roughly constant at the low-mass end. We measure a 1$\sigma$ scatter of about 0.23\,dex in the progenitors of haloes with a $z=0$ mass of $10^9$ M$_\odot$. Noteably, this is similar to the size of the SMHM scatter observed at $z=0$ for massive, star-forming galaxies. This grows to a scatter of 0.5\,dex by $z=0$. In Fig. \ref{fig:scatter-over-time}, we show how the scatter for progenitors of $M_{200}(z=0) = 10^9$ \msun\, haloes increases over time.  We have confirmed that scatter also grows over time following reionisation for low-mass halos in the galaxy formation model \texttt{GALACTICUS} (Ahvazi, private communication).

Lastly, the fact that scatter is not uniform at the low-mass end, but increases with decreasing halo mass is due to the decreasing star formation efficiency of low-mass haloes. In other words, after reionisation, the most massive quenched dwarfs can continue growing in stellar mass through the accretion of smaller galaxies, allowing them to evolve close to the pre-reionisation SMHM track. However, quenched galaxies of lower mass accrete a higher fraction of dark, starless subhaloes, causing them to grow relatively more in their dark versus luminous mass. 

Together, the above results imply that the stellar masses of individual haloes and the scatter in the SMHM relation is predictable if the accretion histories of haloes are known. {\it The stochasticity of star formation is not the main driver of the SMHM scatter} at least down to halo masses of $\sim$10$^9$ \msun.


\subsection{Impact of uncertainties in the physics of reionisation quenching}
\label{sec:smhm:sensitivity}

\begin{figure*}
    \centerline{
    \includegraphics[width=1.04\textwidth]{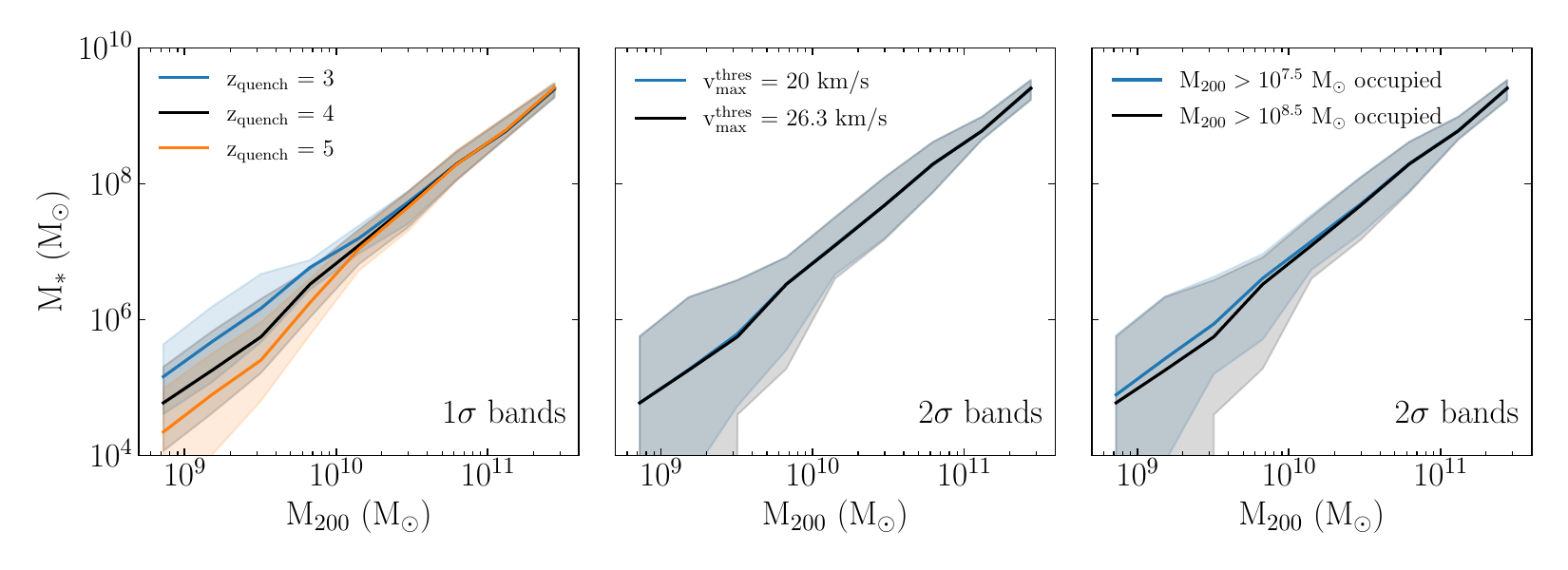}}
    \caption{Impact of uncertainties in the physics of reionisation quenching on the scatter, slope, and shape of the SMHM relation. In all panels, the fiducial model is shown in black. (\emph{Left}) The timing of reionisation quenching affects the shape, slope, and scatter of the SMHM relation at the low-mass end. Earlier quenching at $z_{\rm quench}$ = 5 (orange) results in scatter larger by 0.6\,dex, a steepening of the relation at the low-mass end, and the presence of a ``knee". The bands represent the $1\sigma$ scatter.  Later quenching at $z_{\rm quench} = 3$ (blue) produces smaller scatter and a single power law with no knee. (\emph{Middle}) The threshold $v_{\rm max}^{\rm post}$ required for haloes to reach in order for quenched galaxies to resume star formation after reionisation mildly affects the scatter in the SMHM relation. Decreasing $v_{\rm max}^{\rm post}$ from 26.3 km/s (fiducial; black) to 20 km/s (blue) so that more lower-mass haloes rejuvenate mildly reduces the scatter in the SMHM relation, but does not affect the slope or shape of the relation. Increasing the threshold does not change the scatter. (\emph{Right}) Increasing the occupation fraction such that 50\% of haloes with masses of 10$^{8.5}$ \msun\, instead of 10$^{8.5}$ \msun\, are occupied reduces the steepness of the slope and the scatter in the SMHM relation.  The $1\sigma$ scatter is reduced by $\sim$0.1 dex.}
    \label{fig:sensitivity-smhm}
\end{figure*}

The interplay of reionisation and halo growth histories in shaping the scatter in the SMHM underscores the key role that the physics of reionisation quenching plays in shaping the SMHM relation, including its shape, slope, and normalisation.  In Figure \ref{fig:sensitivity-smhm}, we show how sensitive the location, slope, and scatter in the SMHM relation is to changes in \texttt{DarkLight}'s reionisation quenching model. In all panels, the fiducial relation predicted by \texttt{DarkLight} is plotted in black.

In the leftmost panel, we show the impact of the timing of reionisation quenching, $z_{\rm quench}$. As expected, later quenching at $z_{\rm quench} = 3$ (blue) allows for low-mass haloes to form stars for a longer period of time, elevating the SMHM relation.  While our fiducial model shows signs of a slight steepening in slope for M$_{200} \lesssim 10^{10}$ M$_\odot$, later quenching produces an unbroken power law. We find that if we adopt an earlier quenching redshift, $z_{\rm quench} = 5$ (orange), \texttt{DarkLight} similarly predicts a stronger knee in the SMHM relation. Across the range of quenching redshifts we explored, we find the SMHM slope systematically steepens from $a_{\rm SMHM} = 1.57$ for $z_{\rm quench} = 3$ to $a_{\rm SMHM} = 1.91$ for $z_{\rm quench} = 5$. The scatter at the low-mass end becomes systematically larger with earlier quenching, which increases the fraction of haloes accreted by low-mass galaxies that are devoid of stars. At a halo mass of 10$^9$ M$_\odot$, the 1$\sigma$ scatter increases to $\sim$0.61 dex when we adopt an earlier quenching of $z_{\rm quench}$ = 5.  Conversely, if quenching occurs later, at $z_{\rm quench}$ = 3,  the scatter decreases to 0.50\,dex.  Given that reionisation is believed to occur inhomogenously \citep[e.g.]{Aubert2018, Keating2020, Ocvirk2020, Katz2020}, and that denser regions likely reionise first, this implies that the SMHM may exhibit differences with environment \citep{Christensen24}.

The lower envelope of the SMHM relation is affected by the chosen rejuvenation threshold, i.e. the threshold required for quenched galaxies to resume forming stars. In \texttt{DarkLight}, we adopt a fiducial value of $v_{\rm max}^{\rm post}$ = 26.3 km/s, which best fits the hydrodynamic simulations of the EDGE dwarf galaxies. In the middle column of Fig. \ref{fig:sensitivity-smhm}, we show how the SMHM changes when we adopt $v_{\rm max}^{\rm post}$ = 20 km/s in blue. The shaded bands represent the size of the 2$\sigma$ scatter. The lower threshold allows more galaxies, particularly those in low-mass haloes, to rejuvenate. We see this raise the lower envelope of the SMHM relation $\lesssim 10^{10}$ \msun, decreasing the scatter in the SMHM relation. We find that increasing the threshold to 30 km/s does not change the SMHM nor its scatter.

Another significant uncertainty that affects the evolution of the SMHM after reionisation is the fraction of haloes that are expected to host a galaxy as a function of halo mass, i.e. the occupation function.  Given the role that dark subhaloes play in increasing the size of the scatter in the SMHM (see Sec. \ref{sec:smhm:scatterorigin}), one expects that if a larger fraction of low-mass haloes host galaxies, the smaller the growth in scatter at the low-mass end will be. We find this to be true with \texttt{DarkLight}. In the rightmost panel of Fig. \ref{fig:sensitivity-smhm}, we show that if we adopt an occupation function such that 50\% of halos with masses of M$_{200} > 10^{8.5}$ M$_\odot$ host galaxies, we obtain the relation in black. The shaded bands represent the 2$\sigma$ scatter in the relation. If instead, we shift the occupation function towards lower masses so that 50\% of halos with masses of 10$^{7.5}$ M$_\odot$ host galaxies, we find that the SMHM shifts upwards, as does the bottom of the 2$\sigma$ scatter envelope. The latter threshold is favored by \citet{Nadler2020}, who inferred from the Milky Way dwarfs that nearly all haloes down to 10$^8$ \msun\,host galaxies (see also Sec. \ref{sec:occupationfraction} and Fig. \ref{fig:halo_occupation}). At a halo mass of 10$^9$ M$_\odot$, this reduces the $1\sigma$ scatter by 0.07\,dex, from 0.55\,dex to 0.48\,dex.  

In Table \ref{tab:sensitivity_analysis}, we show how the functional fit to the median and scatter in the SMHM changes for each of the models we explored.

\begin{table}
    \centering
    \caption{Functional fits for SMHM with varied input parameters}
    \begin{tabular}{l l l l l}
         \hline
         variant & $a_{\rm SMHM}$ & $b_{\rm SMHM}$ & $a_{\rm scatter}$ & $b_{\rm scatter}$ \\ \hline
         fiducial & 1.76 & -10.7 & -0.227 & 2.59 \\
         poccALL & 1.72 & -10.3 & -0.194 & 2.22 \\
         poccN18 & 1.73 & -10.4 & -0.202 & 2.3 \\
         $v_{\rm max}^{\rm post}$ = 20 km/s & 1.76 & -10.8 & -0.235 & 2.66 \\
         $v_{\rm max}^{\rm post}$ = 30 km/s & 1.75 & -10.7 & -0.224 & 2.56 \\
         $z_{\rm quench}$ = 3 & 1.57 & -8.71 & -0.217 & 2.45 \\
         $z_{\rm quench}$ = 5 & 1.91 & -12.4 & -0.24 & 2.77 \\ \hline
    \end{tabular}
    \label{tab:sensitivity_analysis}
\end{table}


\section{The halo occupation fraction}\label{sec:occupationfraction}

\begin{figure}
    \centering
    \includegraphics[width=0.5\textwidth]{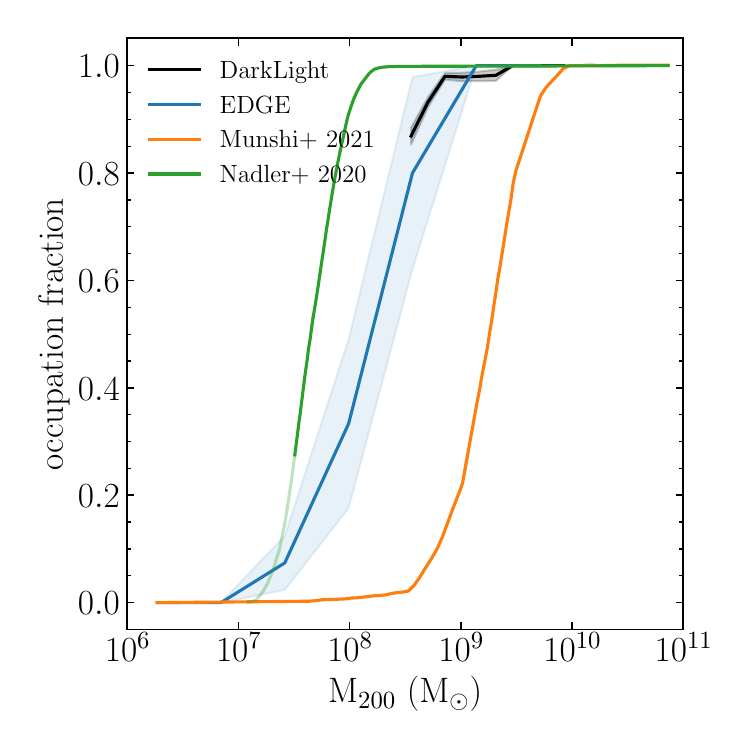}
    \caption{The halo occupation fraction predicted by \texttt{DarkLight} (black), shown down to the resolution limit of the void volume on which it was run (corresponding to a halo mass limit of $\sim$3 $\times 10^8$ \msun).  This is compared to that from the EDGE hydrodynamic zoom simulations (blue), the Marvel-ous Dwarfs and DC Justice League suites \citep{Munshi2021} (orange; based on their sample of galaxies with $>$1 star particle), and inferred from fitting the Milky Way satellite luminosity function by \citep{Nadler2020} (green). The latter is shown in light green at the mass scales below the resolution limit of the simulations based on which the fit was performed.}
    \label{fig:halo_occupation}
\end{figure}

\texttt{DarkLight} also predicts which haloes host galaxies, and which should stay completely dark, i.e. the halo occupation fraction. In Fig. \ref{fig:halo_occupation}, we show \texttt{DarkLight}'s prediction in black. The results are in good agreement with the EDGE simulations, which are shown in blue (bands denote statistical uncertainties), and includes all haloes that have at least one star particle (which are initialized with a mass of 300 \msun, but can undergo mass loss due to supernovae and winds from massive stars).  Due to the resolution of the void volume on which \texttt{DarkLight} was run, we do not have reliable statistics on the occupation fraction of haloes below $M_{200} \sim 3 \times10^8$ \msun~(in contrast, the galaxies in the EDGE suite are zooms that go to much higher resolution). We expect that due to limited numerical resolution, the results from the EDGE simulations are a \emph{lower bound} on the fraction of haloes that host galaxies. We also show the results from the Marvel-ous Dwarfs and DC Justice League simulations \citep{Munshi2021}, again showing all haloes with at least one star particle, and from a fit to observed Milky Way satellite galaxies \citep{Nadler2020}. \texttt{DarkLight} and EDGE predict haloes are occupied to lower halo masses than in \citet{Munshi2021}, but higher halo masses than inferred by \citet{Nadler2020}.  \citet{Munshi2021} show that the occupation fraction inferred is strongly dependent on resolution, and thus the mismatch between \citet{Nadler2020} versus that from \texttt{DarkLight} and both simulation suites could be due to issues in resolution.

\section{Star-forming versus quenched dwarfs}
\label{sec:starform}

\begin{figure}
    \includegraphics[width=0.5\textwidth]{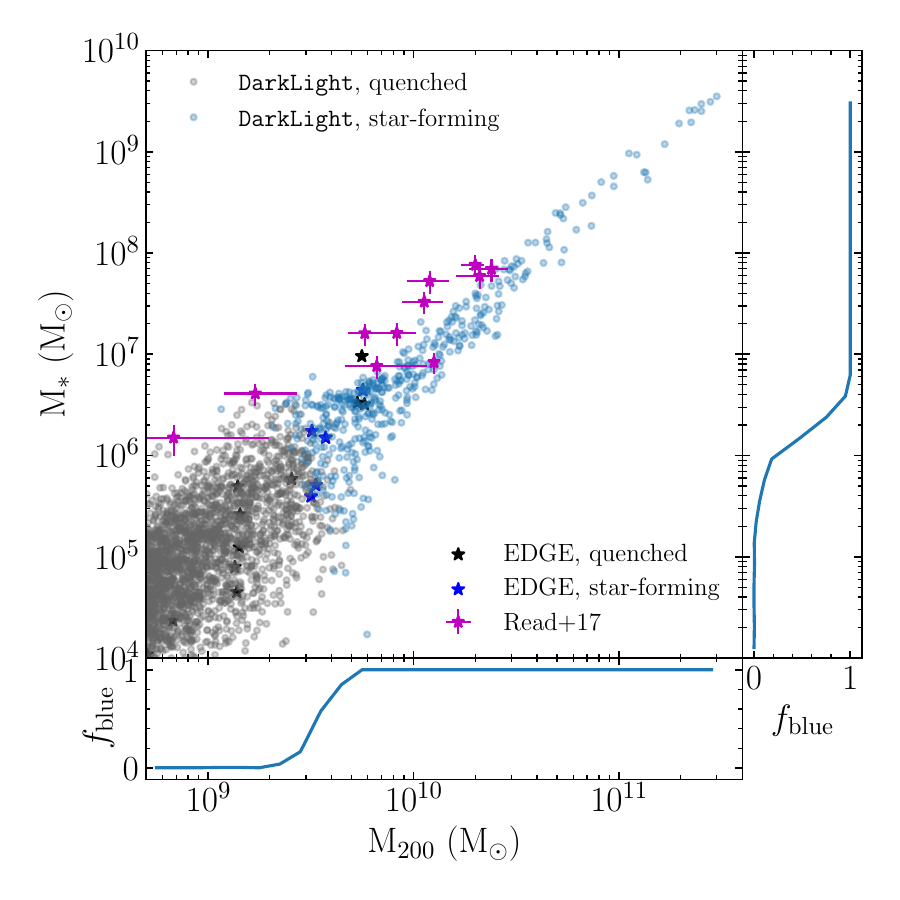}
    \caption{Distribution of star-forming and quiescent dwarf galaxies predicted by \texttt{DarkLight}. The points and symbols are as in \ref{fig:smhm}, but with galaxies shaded blue if they are identified to be star-forming today, which we take to be haloes with $v_{\rm max}(z=0) > v_{\rm max}^{\rm post}$ = 26.3 km/s, or shaded gray if they are quiescent.  Similarly, dwarfs in the EDGE simulations suite (stars) are shaded blue if they are star-forming at the present day.  Observed gas-rich, star forming, galaxies are plotted in magenta \citep{Read2017}. The bottom and right panels show the fraction of dwarfs that are star-forming today, as a function of halo or stellar mass, respectively. There is a sharp transition from quiescent to star-forming at halo masses of a few 10$^9$ \msun, and stellar masses of a few 10$^6$ \msun. \texttt{DarkLight} currently does not model self-quenching due to starbursts, which have been observed in some EDGE dwarfs (black stars at $M_{200} \sim 10^{10}$ \msun; \citealt{gray24}). As such, the transition between quenched and star-forming galaxies may not be quite as steep once these are factored in, and the boundary we find between them should be taken as a theoretical lower bound.}
    \label{fig:fblue}
\end{figure}

The presence of star-forming (or recently star-forming) dwarf galaxies with very low stellar masses has been an enduring puzzle. These include galaxies such as Aquarius, CVndwIa, Leo T, Leo P, Antila B, and Peg W, which have stellar masses as low as $\sim$10$^5$ \msun, and are thought to inhabit haloes of mass less than $\sim$5 $\times $10$^9$ \msun\, \citep{Read2017,Zoutendijk21}. While recent high-resolution simulations \citep{Wright2019, Rey2020, Applebaum2021, Gutcke2022} and some semi-empirical models \citep{Benitez-Llambay2021, OLeary2023, Ahvazi2024} have begun to reproduce analogs of galaxies like Leo T, including their HI content \citep{Rey2022}, dwarfs such as Aquarius and CVndwIa, that have high stellar mass at such low halo mass, remain a challenge to reproduce.

As \texttt{DarkLight} produces star formation histories, it can identify which haloes are likely star-forming or quenched today. We can thus search for analogs of low-mass, star-forming dwarfs. In Fig. \ref{fig:fblue}, we plot galaxies predicted by \texttt{DarkLight}, but shade in light blue those haloes predicted to be star-forming today (i.e. $v_{\rm max}(z=0) >$ 23.5 km/s). We find that haloes with $M_{200} \gtrsim 5 \times 10^9$ \msun~ are all star-forming. Below this mass, some haloes without ongoing star formation begin to appear, increasing in number until $M_{200} \lesssim 2 \times 10^9$ \msun, below which all haloes are quiescent.  This agrees with the threshold found in other theoretical work \citep{Christensen24}. The fraction of haloes that are star-forming as a function of halo mass is shown in the bottom panel. We find that 50\% of dwarfs are star-forming at a halo mass of $\sim 4 \times 10^9$ \msun.

In terms of stellar mass, all galaxies are star forming down to $M_* \sim 5 \times 10^6$ \msun, at which point quiescent dwarfs begin to appear. The fraction of galaxies that are star-forming as a function of stellar mass is shown in the right panel. About 50\% of dwarfs are star-forming at $M_* \sim 2 \times 10^6$ \msun.  There is a small but significant tail of star-forming galaxies with $M_* \lesssim 10^6$ \msun. 

We note that \texttt{DarkLight} does not include self-quenching of dwarfs due to starbursts, and thus some dwarfs that we predict to be star-forming may be quenched (e.g. black stars at $M_{200} \sim 10^{10}$ \msun~in Fig. \ref{fig:fblue}; and see also \citealt{gray24}). The mass thresholds above can thus be considered as \emph{lower limits} on the mass scale at which dwarf galaxies are star forming. Further, the transition between star-forming and quiescent dwarfs may not be quite as steep as we predict.

With this in mind, we find that \texttt{DarkLight} predicts star-forming analogs of Aquarius and CVndwIa, which are represented by the two leftmost magenta data points, respectively. The haloes that correspond to these particularly low-mass dwarf irregulars tend to be early formers that assembled their stellar content early, then resumed forming stars only in the last few billion years. Although they are rare, \texttt{DarkLight} also predicts analogs for dwarfs with even lower stellar masses such as Leo T, Leo P, Antila B, and Peg W.  In contrast to Aquarius, these low stellar mass dwarfs tend to be late formers that have undergone recent mass growth. A detailed comparison of predictions by \texttt{DarkLight} to observational measurements of the star formation histories and kinematics of these low-mass galaxies will be presented in future work. These lowest-mass star-forming galaxies are the tip of the iceberg of a much larger population of quiescent galaxies at this mass scale. Indeed, a quenched field dwarf that may be consistent with reionisation quenching has recently been discovered \citep[Hedgehog,][]{Li2024}.  Such galaxies represent highly valuable constraints on galaxy formation models, which must reproduce these systems. We will further test this in future work by simulating these low-mass star-forming analogs with full hydrodynamics.


\subsection{Impact of uncertainties in the physics of reionisation quenching}
\label{appendix:sensitivity:blue}

\begin{figure*}
    \centering
    \centerline{}
    \includegraphics[width=0.49\textwidth]{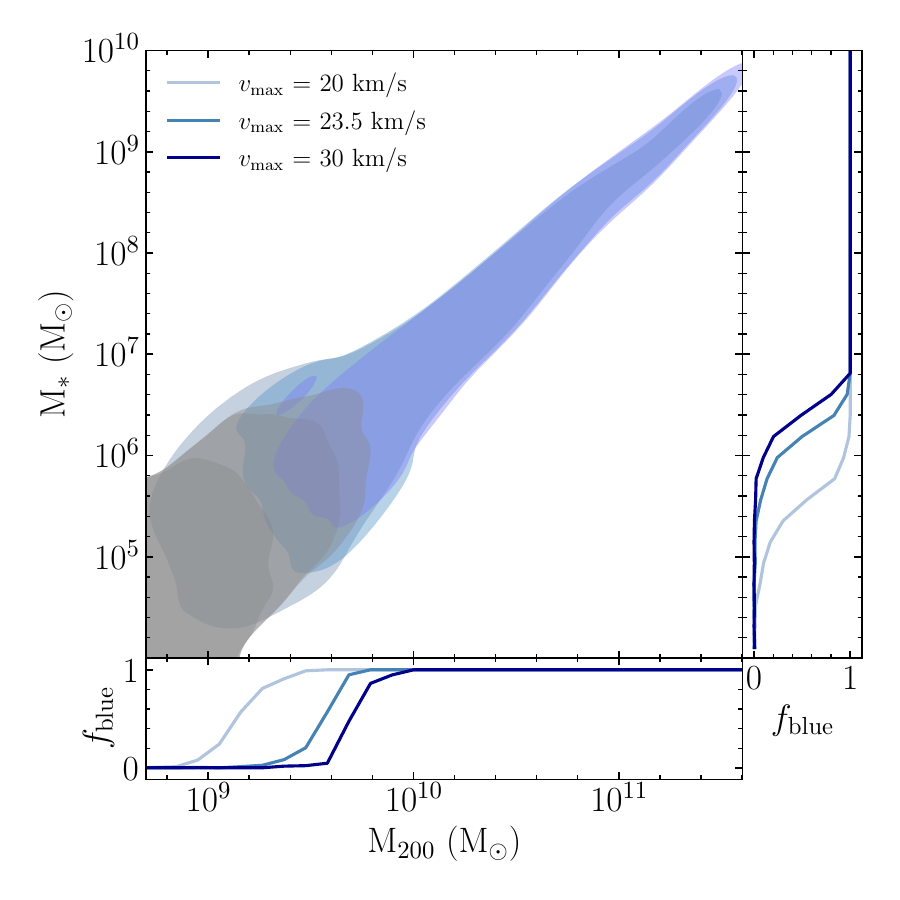}
    \includegraphics[width=0.49\textwidth]{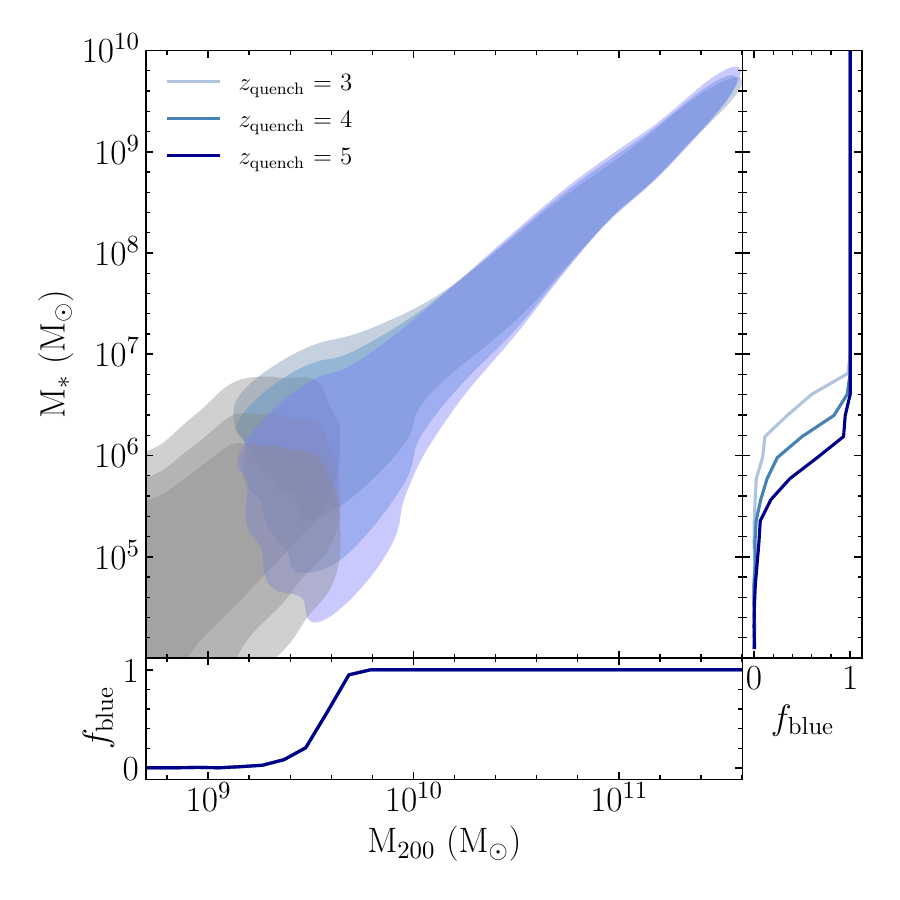}
    \caption{Change in the distribution of star-forming (blue) versus quenched (gray) dwarf galaxies due to modifications to \texttt{DarkLight}'s input physics.  \emph{Left}: Lowering (raising) the threshold needed for dwarf galaxies quenched by reionisation to rejuvenate shifts the boundary between star-forming dwarfs and reionisation relics to lower (higher) masses. \emph{Right}: If galaxies quench earlier (later), they form fewer (more) stars, pushing the boundary between star-formers and reionisation relics to lower (higher) stellar masses. The boundary in halo mass remains unchanged.}
    \label{fig:sensitivity-blue}
\end{figure*}

Here, we show how sensitive the boundary between star-forming dwarfs and reionisation relics is to uncertainties in the physics of reionisation quenching.  In the left panel of Fig. \ref{fig:sensitivity-blue}, we show how changes in the threshold $v_{\rm max}$ required to for galaxies quenched by reionisation to reach to resume star formation affects the boundary. As expected, lowering the threshold lowers the (halo and stellar) mass of the boundary between star-forming dwarfs and reionisation relics, and vice versa. Lowering the threshold from the fiducial value of $v_{\max}^{\rm post}$ = 23.5 km/s to 20 km/s has a dramatically larger effect than increasing it to 30 km/s. A threshold of 20 km/s lowers the stellar mass at which 50\% of galaxies are star-forming by nearly an order of magnitude to $3 \times 10^5$ \msun, compared to the fiducial value of $2 \times 10^6$ \msun, while increasing it to 30 km/s raises it to $3 \times 10^6$ \msun. In terms of halo mass, a threshold of 20\,km/s lowers the 50\% point to $10^9$ \msun, compared to the fiducial value of $4 \times 10^9$ \msun, while increasing the threshold to 30 km/s raises it to $5 \times 10^9$ \msun.

In the right panel of Fig. \ref{fig:sensitivity-blue}, we change the redshift at which dwarf galaxies quench. This does not alter the halo mass of the boundary between star-forming and quenched dwarfs. However, as earlier quenching truncates stellar mass growth earlier, their stellar masses are lower than if quenching occurred later. This shifts the boundary between star formers and quenched dwarfs to lower stellar masses. For a quenching redshift $z_{\rm quench} = 5$, the stellar mass at which 50\% of galaxies are star-forming shifts to $M_* = 7 \times 10^5$ \msun, from the fiducial value of $2 \times 10^6$ \msun, while later quenching at $z_{\rm quench} = 3$ shifts the threshold to $4 \times 10^6$ \msun.


\section{Discussion}
\label{sec:discussion}

\subsection{Comparison with other galaxy formation models}
\label{sec:smhm:SAMs}

\begin{figure}
    \includegraphics[width=0.5\textwidth]{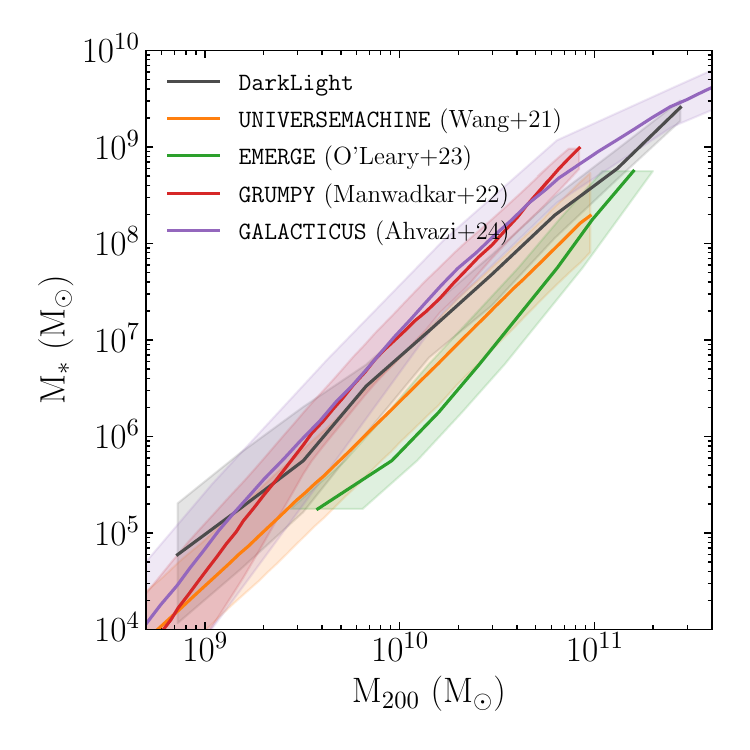}
    \caption{SMHMs predicted by various galaxy formation models. Bands denote 1$\sigma$ contours. \texttt{DarkLight}, \texttt{UNIVERSEMACHINE} \citep{Wang2021}, and \texttt{EMERGE} \citep{OLeary2023} are semi-empirical models, \texttt{GRUMPY} \citep{Manwadkar2022} is a regulator model, and \texttt{GALACTICUS} \citep{Ahvazi2024} is a fully semi-analytic model. \texttt{UNIVERSEMACHINE} currently does not include reionisation. For \texttt{EMERGE}, the SMHM based on the ``High-$z$ quench" reionisation model is shown . Note that the contours for \texttt{EMERGE} denote the scatter in halo mass for a given stellar mass bin rather than the scatter in stellar mass for a given halo mass, in contrast to the other relations.  \texttt{UNIVERSEMACHINE} and \texttt{SatGen} do not model reionisation.}
    \label{fig:smhm-SAMs}
\end{figure}

Below scales of $\sim$10$^{10}$ \msun, there are a handful of galaxy formation models tailored to model dwarf galaxies.  Those calibrated on higher-mass galaxies do not necessarily work at low-mass scales, in part due to the lack of reionisation quenching and its impact on galaxy growth. Those that do exist adopt different approaches to modelling dwarf evolution, ranging from \texttt{UNIVERSEMACHINE} \citep{Behroozi2019} and \texttt{EMERGE} \citep{Moster2018}, semi-empirical models constrained by observations, \texttt{GRUMPY}, a regulator model that models gas inflow, outflow, and star formation \citep{Kravtsov2022}, \texttt{Galacticus} and \texttt{GALFORM}, fully semi-analytic codes \citep{Benson2012, Cole1994}, and \texttt{SatGen}, an emulator based on numerical simulations \citep{Jiang2021}.

\subsubsection{\emph{\texttt{UNIVERSEMACHINE}}}

\texttt{DarkLight}'s reliance on empirical relations to calibrate the dependence of star formation rates on $v_{\rm max}$ is most similar to \texttt{UNIVERSEMACHINE} \citep{Behroozi2019}. However, they differ in two important regards.  Firstly, \texttt{DarkLight} is calibrated on observations where available, and supplemented by high-resolution simulations where they run out, and is thus calibrated to $M_{200} \sim 10^9$ \msun, corresponding to a median $M_* \sim 10^5$ \msun). In contrast, \texttt{UNIVERSEMACHINE} is exclusively fit to observations.  While \texttt{DarkLight} fits mean SFRs to fit two $\langle$SFR$\rangle$-$v_{\rm max}$ relations, \texttt{UNIVERSEMACHINE} performs a MCMC fit for the instanteous SFR as a function of redshift using a wide array of observations concerning massive galaxies, including stellar mass and UV luminosity functions, specific and cosmic SFRs, and quenched fraction, among others.  Recently, it was recalibrated with data from the SAGA survey of satellites around MW analogues, which is largely complete down to dwarfs with $M_* \gtrsim 10^{7.5}$ \msun~\citep{Mao2024}, and isolated SDSS field galaxies down to $M_* \gtrsim 10^{7}$ \msun, extending its calibration range \citep{Wang2024}. Although \texttt{DarkLight} is calibrated down to much lower mass scales, its dependence on hydrodynamic simulations for $v_{\rm max} \lesssim$ 20 km/s implies that our results at the low-mass end are dependent on the subgrid physics adopted in EDGE. However, the EDGE simulations match observed dwarfs well \citep[e.g.][]{Agertz2020, Richstein2024}.

A second important difference is that we model reionisation, while \texttt{UNIVERSEMACHINE} does not. This causes \texttt{UNIVERSEMACHINE} to predict too many star-forming dwarf galaxies \citep[][although this has improved with the addition of environmental quenching, c.f. \citealt{Wang2024}]{Wang2021}.  \texttt{DarkLight} predicts a more realistic distribution of star-forming and quenched dwarfs (c.f. Sec \ref{sec:starform}).  Despite these differences, \texttt{DarkLight} and \texttt{UNIVERSEMACHINE} produce similar SMHMs, as shown in Fig. \ref{fig:smhm-SAMs}, albeit systematically lower in stellar mass by about a factor of 2-3, and reproduces the SMHM inferred from the MW dwarfs \citep{Nadler2020} well.  As \citet{Wang2021} predicted, \texttt{UNIVERSEMACHINE} produces realistic stellar masses as the suppression of star formation due to reionisation quenching is made up for by enhanced SFRs before reionisation, which is observed in the EDGE simulations. However, as they do not include reionisation quenching, the SMHM relation predicted by \texttt{UNIVERSEMACHINE} does not exhibit an increase in scatter at low masses.

Lastly, we note that we have calibrated on isolated dwarfs, while \texttt{UNIVERSEMACHINE} includes environmental physics \citep{Wang2024}.  We will add environmental physics in future work.

\subsubsection{\emph{\texttt{EMERGE}}}

\texttt{EMERGE} is also based on a wide range of empirical relations, initially primarily of massive galaxies \citep{Moster2018}.  However, instead of relying on a relation between SFR and $v_{\rm max}$, \texttt{EMERGE} scales SFRs based on the change in mass of a dark matter halo over time. This assumes that accreted gas can be converted into stars within a dynamical time.  This assumption breaks down at dwarf scales due to reionisation quenching. \texttt{EMERGE} was recently updated to include reionisation quenching, fitting to a stellar mass function extended to include observed Local Group dwarfs with $M_* \gtrsim 10^5$ \msun\, to infer the threshold mass below which galaxies were quenched \citep{OLeary2023}.  Their preferred model for comparison (`logistic', see their Fig. 2) quenches dwarfs around $z \sim 4$ with masses $M_{\rm halo} \lesssim 2 \times 10^9$ \msun\, (note that while they explored the redshift dependence of the mass cutoff, they did not find it was strongly constrained by the data).  In contrast, \texttt{DarkLight} identifies quenched galaxies as those with $v_{\rm max} <$ 26.3 km/s at $z_{\rm quench} = 4$, which corresponds to a similar cutoff halo mass as in \texttt{EMERGE}.  In Fig. \ref{fig:smhm-SAMs}, we show the SMHM of \texttt{EMERGE} for their model with reionisation quenching in green. Unlike the other SMHM relations, the median and scatter were reported for galaxies binned in stellar mass, rather than halo mass.  With this in mind, we find that the median relation predicts stellar masses an order of magnitude lower than \texttt{DarkLight}. We note that their reference model (not plotted), which does not include reionisation quenching, produces a SMHM relation and scatter more consistent with, though still systematically lower than, that predicted by \texttt{DarkLight} and other galaxy-halo models.

\subsubsection{\emph{\texttt{GRUMPY}}}

\texttt{GRUMPY} is a galaxy formation code tailored to dwarf galaxies that distills key galaxy formation physics into a few differential equations that describe gas inflow, outflow, and sinks---an approach described as a ``regulator" or ``bathtub" model \citep[e.g.][]{peng10}.  Like \texttt{EMERGE}, they tie gas inflow rates to the dark matter accretion rate, which are derived from dark matter only simulations. However, they include more detailed modeling of reionisation quenching, H$_2$ cooling, and galactic outflows, among others \citep{Kravtsov2022}. \texttt{GRUMPY}'s prediction for the SMHM relation, based on forward-modelling the MW satellites \citep{Manwadkar2022}, is shown in red in Fig. \ref{fig:smhm-SAMs}. This favors a broad range of reionisation redshifts spanning $z \sim 7-10$. Like \texttt{DarkLight}, they also find an increase in scatter at halo masses below $10^{10}$ \msun. \citet{Kravtsov2022} find that adopting earlier reionisation results in a similar steepening in the slope at the low-mass end. They also find that the SMHM relation is sensitive to the wind model they adopt. We note that the EDGE simulations upon which \texttt{DarkLight} is based self-consistently resolves momentum and energy injection by individual supernovae, but does not include a model for radiative transfer. The latter acts to suppress strong outflows in the lowest mass dwarfs \citep{Agertz2020}, lowering the stellar mass for a given halo mass.

\subsubsection{Semi-analytic models: \emph{\texttt{Galacticus}}}

Semi-analytic codes include detailed models of the many physical processes involved in galaxy formation. While a number of semi-analytic codes exist, \texttt{Galacticus} in particular has recently been extended to dwarf scales by \citet{Ahvazi2024}. They include an updated UV background model for reionisation (favoring a reionisation redshift of $z = 7.8$), a new model for the metallicity of the intergalactic medium and, crucially, H$_2$ cooling, which they find is required to match the mass-metallicity relation and the inferred halo occupation function of the MW satellites.  \texttt{Galacticus}'s predictions for the SMHM is shown in Fig. \ref{fig:smhm-SAMs} in purple.  It closely matches the prediction by \texttt{DarkLight} and exhibits an increase in scatter at the low mass end that similarly grows following reionisation quenching (Ahavazi, private comm.). It exhibits a steepening of the SMHM slope at slightly higher halo masses of $\sim$10$^{11}$ \msun\, than in other work \citep[e.g.][]{Munshi2021}.

While \texttt{GALFORM}, another semi-analytic code, has also been used in studies of dwarf galaxies, including to study the impact of reionisation on the luminosity function at the faint end \citep[e.g.][]{Bose2018}, their predictions for the SMHM relation have not been published, so we do not discuss this model in detail here.

\subsubsection{\emph{\texttt{SatGen}} and other models that sample from a SMHM relation}

While the models discussed above \emph{predict} the SMHM relation from various assumptions on how to model galaxy formation, many dwarf galaxy studies simply adopt a published SMHM relation and sample stellar masses from its scatter. Of note, this is the approach taken by \texttt{SatGen} \citep{Jiang2021}, an emulator based on numerical simulations that tracks in detail the evolution of satellites in a host halo. \texttt{SatGen} utilises a SMHM relation derived via abundance matching by \citet{Rodriguez-Puebla2017}, which is calibrated for haloes with present-day masses $M_{\rm vir} \gtrsim 10^{10}$ \msun, and thus relies on extrapolations to dwarf galaxies below this mass (i.e. likely for most of the ultrafaint dwarf galaxies).  Notably, it adopts a constant scatter of 0.16\,dex, and thus does not include the impact of reionisation on the shape, slope, normalisation, or scatter in the SMHM relation. Further, we note that studies that similarly assign haloes a stellar mass by simply sampling from the scatter of a SMHM relation will miss the dependence of stellar masses on dark matter accretion histories.


\subsection{Comparison with simulations}
\label{sec:smhm:sims}

In light of the sensitivity of the SMHM relation to the halo occupation function and uncertainties in the physics of reionisation quenching, it is not surprising that hydrodynamic simulations differ in their predictions for the SMHM by as much as two orders of magnitude below halo masses of $\sim$10$^{10}$ \msun\,\citep[see Fig. \ref{fig:smhm} and][]{Garrison-Kimmel2017}.  These physics are highly uncertain in simulations of dwarf galaxies. 

Resolution is important to avoid numerical quenching due to poorly resolved gas content and star formation in low-mass dwarf galaxies. \citet{Munshi2021} showed that the halo occupation function can shift by an order of magnitude depending on one's resolution.  Recent simulations have begun to achieve sufficient resolution to resolve $\lesssim 10^{10}$ \msun\,haloes.  Resolution studies performed by some groups running very high-resolution hydrodynamic simulations show that stellar masses of $10^9$ \msun\, haloes remain largely unchanged with increased resolution \citep[e.g.][]{Munshi2021}, while others find that increasing their dark matter resolution by a factor of $\sim$10 caused their stellar masses to increase by a factor of 2 \citep{Agertz2020}.  There are still significant differences between these simulation suites, underscoring that resolution cannot explain the entirety of the differences between hydrodynamic simulations.

The remaining discrepancies likely owe in large part to differing subgrid physics governing star formation and evolution.  Importantly, supernovae feedback models are highly uncertain yet are key to determining stellar masses of dwarf galaxies.  \citet{Agertz2020} showed that a given halo can form more than an order of magnitude more stars if it has weak or no feedback; radiative transfer can further reduce the stellar mass by another order of magnitude.  Non-equilibrium cooling can then counteract this by allowing more stars to form (Rey et al. in prep 2024).  \citet{Munshi2019} showed that using a classical temperature-density threshold for star formation produced more dwarf galaxies than using a non-equilibrium H$_2$-cooling threshold.  Furthermore, few simulations model cosmic ray feedback and magnetic fields, which affect the stellar mass of dwarf galaxies in non-linear ways \citep[e.g.][]{Hopkins2020, Martin-Alvarez2023}.  However, subgrid physics is not the only driver of differences in stellar masses. \citet{Hu2023} showed that Eulerian and Lagrangian codes---all sub-grid physics being as equal as possible---can produce significant differences in the burstiness of star formation, which can impact the stellar mass they form.

The fact that simulations begin to differ at $10^{10}$ \msun\,is perhaps key to note.  This is the scale at which reionisation relics begin to dominate the population (e.g. \S\ref{sec:starform}).  Thus the stellar content of these haloes will largely (if not entirely) be set by the physics of reionisation and of galaxy formation in the early universe.  \citet{Benitez-Llambay2020} showed that both can shift the halo occupation function by orders of magnitude.  Further complicating matters, reionisation is believed to be inhomogenous, with dense environments reionising earlier than low-density regions \citep[e.g.]{Aubert2018, Keating2020, Ocvirk2020, Katz2020}, implying that the SMHM and its scatter should vary with environment \citep[as shown by][]{Christensen24}.  H$_2$ cooling appears to be an essential ingredient to reproduce reionisation relics around the MW \citep{Manwadkar2022, Ahvazi2024}.  Reducing uncertainties on the reionisation models and the subgrid physics that governs the formation of dwarf galaxies before reionisation will be key to understanding the true form of the SMHM.


\subsection{Limitations of \texttt{DarkLight}} 
\label{sec:disc:improvements}

In this section, we briefly highlight a few current known caveats and limitations of {\tt DarkLight} and discuss where and how these can be addressed in future work.

\begin{enumerate}[labelwidth=0em, labelsep=0.5em, leftmargin=\parindent, itemsep=2pt]

\item \textbf{Episodic star formation}: \texttt{DarkLight} assumes a \emph{mean} star formation rate whenever haloes are identified to be actively forming stars and does not model stochasticity. However, observed star forming dwarf irregulars \citep{Collins22}, as well as EDGE dwarfs that become sufficiently massive to reignite star formation after reionisation, exhibit episodic star formation (Fig. \ref{fig:validation}). We will explore adding perodicity and/or stochasticity to the mean star formation rates in \texttt{DarkLight}, which can be calibrated both by simulations like EDGE, or empirically fit to the latest data constraints \citep[e.g.][]{Collins22}.

\item \textbf{Starbursts and self-quenching}: \texttt{DarkLight} can produce a temporary burst of star formation in dwarfs quenched by reionisation due to mergers (e.g. Halo 600 in Fig. \ref{fig:validation}) caused by temporary rises in $v_{\rm max}$ that accompany mergers, if it passes above the rejuvenation threshold. However, in EDGE, we find that such mergers can excite starbursts with star formation rates significantly higher than given by our $\langle$SFR$\rangle$-$v_{\rm max}$ relation, with the feedback from such events being sufficient to fully self-quench the dwarf \citep{gray24}.  We will explore adding such a starburst model to \texttt{DarkLight} in future work (Gray et al., in prep).

\item \textbf{Calibration on EDGE}: For low mass dwarfs, \texttt{DarkLight} is currently calibrated on the EDGE simulation suite, and thus its predictions are as accurate insofar as these simulations are a reliable facimile of nature.  As discussed in Sec. \ref{sec:smhm:sims}, switching to an alternative cosmological code and/or different subgrid model choices could yield significant changes. \texttt{DarkLight} can be calibrated on other simulations, and ultimately be calibrated fully on data, if data of sufficient quality become available for galaxies with $v_{\rm max} < 20$\,km/s.  We are exploring how changes in subgrid physics in the EDGE2 suite \citep[][Rey et al, in prep.]{Muni2024} modify our predictions for the SMHM relation.

\end{enumerate}


\section{Conclusions}
\label{sec:conclusions}

We have presented a new semi-empirical code, \texttt{DarkLight}, designed to predict the stellar mass assembly history of dwarf galaxies, down to the very faintest systems. We found that the interplay between reionisation and halo growth histories is key to determining the slope, scatter, and shape of the stellar-mass--halo-mass (SMHM) relation. Our main conclusions are as follows:

\begin{itemize}[labelwidth=0em, labelsep=0.5em, leftmargin=\parindent, itemsep=2pt]

\item For higher-mass dwarfs with $M_{200} \gtrsim 10^{10}$ \msun, \texttt{DarkLight}'s prediction for the SMHM relation agrees well with abundance matching, simulations, and observational studies. Below this scale, where studies differ significantly, \texttt{DarkLight} predicts a large scatter, with a 1$\sigma$ symmetric scatter at $M_{200} = 10^9$ \msun~of 0.5\,dex.

\item The scatter in the SMHM relation down to $M_{200} \lesssim 10^9$ \msun~is not predominantly due to stochastic star formation. Instead, the truncation of galaxy growth due to reionisation, which decouples the growth of the stellar mass from the halo mass of a galaxy, combined with the wide range of halo assembly histories, accounts for the majority of the scatter.

\item The scatter in the SMHM is constant for low-mass haloes at reionisation, but grows after reionisation quenching takes place. Like previous simulations-based results, we find the size of the scatter increases with decreasing halo mass, beginning at the mass scale where reionisation quenches low-mass dwarfs. This is due to the increasing dominance of dark haloes (i.e. those devoid of stars) accreted by haloes of decreasing mass. The scatter at the high mass end, where reionisation does not quench galaxies, does not evolve significantly. 

\item While we do not find a significant break in the slope of the SMHM, one can be introduced if we push reionisation quenching to earlier times than in our default model. The break occurs at the boundary where reionisation begins to quench low-mass dwarfs.  If a knee exists, the location of the knee, as well as the slopes above and below it, could help constrain the timing of reionisation and the mass scale at which dwarf galaxies are quenched. 

\item The two orders of magnitude difference between the location, slope, and scatter in the SMHM relation predicted by different hydrodynamical simulations occurs at the mass scale of reionisation relics. Reducing uncertainties on the subgrid physics that govern the formation of dwarf galaxies before reionisation, and uncertainties in the reionisation model, is likely key to understanding the true form of the SMHM. 

\end{itemize}

\noindent
Models that have explored the SMHM relation at lower masses have found intriguing signs of a flattening, likely a signature of H$_2$ cooling \citep[e.g.][]{Manwadkar2022, Ahvazi2024}.  While we currently do not have the resolution to model galaxies down to this regime, our $\langle$SFR$\rangle$-$v_{\rm max}$ relation appears to exhibit a flattening for small haloes, which may reproduce the trend in stellar masses. We are currently in the process of running a higher resolution suite of haloes at lower mass scales, which will enable us to study the SMHM relation in the ultra-low mass regime.

Assembly histories---and thus the SMHM relation---depend on the background cosmology.  Alternative dark matter models such as warm dark matter predict different assembly history statistics \citep[e.g.][]{Ludlow2016}, which should then map onto systematic differences in the SMHM. We are currently testing whether \texttt{DarkLight} needs re-calibration to be applied to $\Lambda$WDM cosmologies.

Stellar mass is not the only property affected by the assembly history of a galaxy. Later assembly may produce lower stellar masses in addition to larger galaxy sizes and lower metallicities \citep{Rey2019}. As such, the scatter in each of these properties may not be fully independent, and galaxy formation models that assign galaxy properties treating each the scatter in each of these properties as independent may miss correlations that could be harnessed to improve dark matter constraints from up-coming surveys. We are currently implementing the dependence on assembly histories for galaxy sizes (Nigudar et al., in prep) and HI content (Hutton et al., in prep) into \texttt{DarkLight}, and plan to expand it to include also metallicities, kinematics, and shapes.


\section*{Acknowledgements}

SYK thanks Annika Peter, Niusha Ahvazi, Andrew Benson, Ethan Nadler, Yao-Yuan Mao, and Marla Geha for helpful discussions that improved this manuscript.  JIR would like to acknowledge support from STFC grants ST/Y002865/1 and ST/Y002857/1.  MR is supported by the Beecroft Fellowship funded by Adrian Beecroft.  MO acknowledges funding from the European Research Council (ERC) under the European Union’s Horizon 2020 research and innovation programme (grant agreement No. 852839).  ET acknowledges the UKRI Science and Technology Facilities Council (STFC) for support (grant ST/V50712X/1).  OA acknowledges support from the Knut and Alice Wallenberg Foundation, the Swedish Research Council (grant 2019-04659), and the Swedish National Space Agency (SNSA Dnr 2023-00164).  This work was performed using the DiRAC Data Intensive service at Leicester, operated by the University of Leicester IT Services, which forms part of the STFC DiRAC HPC Facility (\href{www.dirac.ac.uk}{www.dirac.ac.uk}).  The authors acknowledge the use of the University of Surrey Eureka supercomputer.

\emph{Software}:  \texttt{pynbody} \citep{Pontzen2013}, \texttt{tangos} \citep{Pontzen2018}, \texttt{colossus} \citep{Diemer2018}, SciPy \citep{Jones2001}, Matplotlib \citep{Hunter2007}, NumPy \citep{vanderWalt2011},  Jupyter (\href{jupyter.org}{jupyter.org}).


\section*{Author Contributions}

Contributions from the authors were, based on the \href{https://credit.niso.org/}{CRediT (Contribution Roles Taxonomy)} system: 
\textbf{SK}: conceptualization; investigation; methodology; software; formal analysis; visualisation; writing - original draft, review \& editing. 
\textbf{JR}: conceptualization; methodology; validation and interpretation; writing - original draft, review \& editing; funding acquisition; project administration.  
\textbf{MR}: data curation; software; resources; writing - review \& editing. 
\textbf{MO}: data curation; software; formal analysis; writing - review \& editing.
\textbf{SN}: investigation; methodlogy; formal analysis; software.
\textbf{AP}: data curation; software; writing - review \& editing.
\textbf{ET}: data curation.
\textbf{OA}: data curation; software; writing - review \& editing.
\textbf{PD}: writing - review \& editing.


\appendix

\section{Correlations with SFR}
\label{appendix:M200}

\begin{figure}
    \centering
    \includegraphics[width=0.5\textwidth]{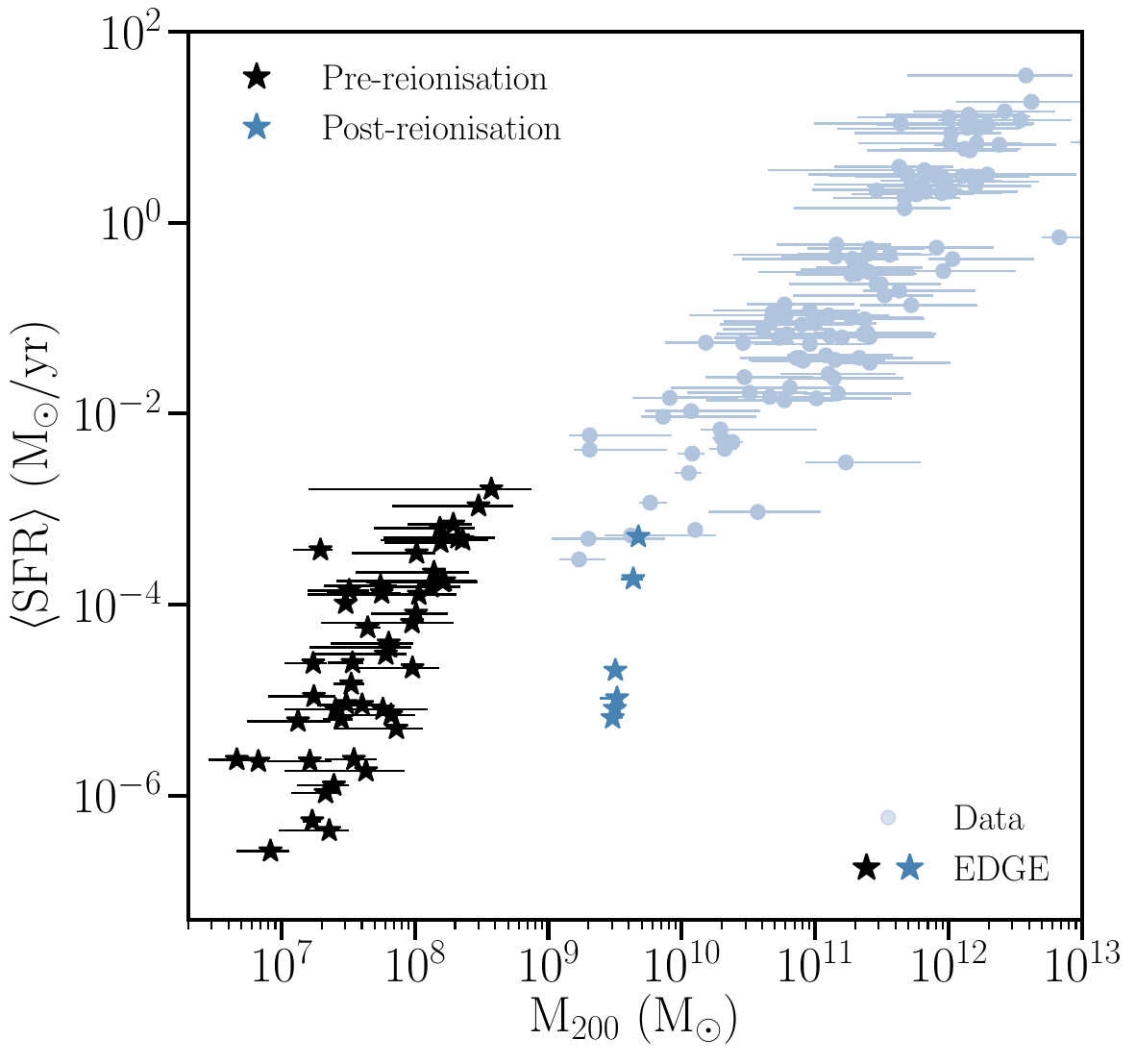}
    \caption{Relation between the average star formation rate and halo mass, $M_{200}$.  There is significantly more scatter and larger uncertainties in the data than with $v_{\rm max}$ (c.f. Fig. \ref{fig:sfr-vmax}).  We thus use the $\langle$SFR$\rangle$-$v_{\rm max}$ relation in \texttt{DarkLight}.}
    \label{fig:sfr-M200}
\end{figure}

In \citet{Read2017}, a tight correlation was found between the average star formation rate and the halo mass $\langle$SFR$\rangle$-$M_{200}$ for Local Group dwarf galaxies.  We test whether this holds for simulated galaxies in EDGE and a larger sample of observed, nearby isolated galaxies (see Sec. \ref{sec:darklight:sfr-vmax} for details on these datasets).  In Fig. \ref{fig:sfr-M200}, we show the correlation between the mean star formation rate, $\langle$SFR$\rangle$ and halo mass, $M_{200}$.  As in Fig. \ref{fig:sfr-vmax}, we show pre-reionisation values in black and post-reionisation values in blue, and data from simulations with stars and from data in circles.  While a correlation between $\langle$SFR$\rangle$ and $M_{200}$ does exist, it has significantly larger scatter and uncertanties than it does with the peak rotation speed, $v_{\rm max}$ (see Figure \ref{fig:sfr-M200}). For this reason, we use instead the $\langle$SFR$\rangle$-$v_{\rm max}$ relation in \texttt{DarkLight}.

\section{$v_{\rm max}$ in DMO vs baryonic simulations}
\label{appendix:vmaxDMO} 

\begin{figure}
    \centering
\includegraphics[width=0.5\textwidth,clip=true,trim=0in 0.5in 0in 1in]{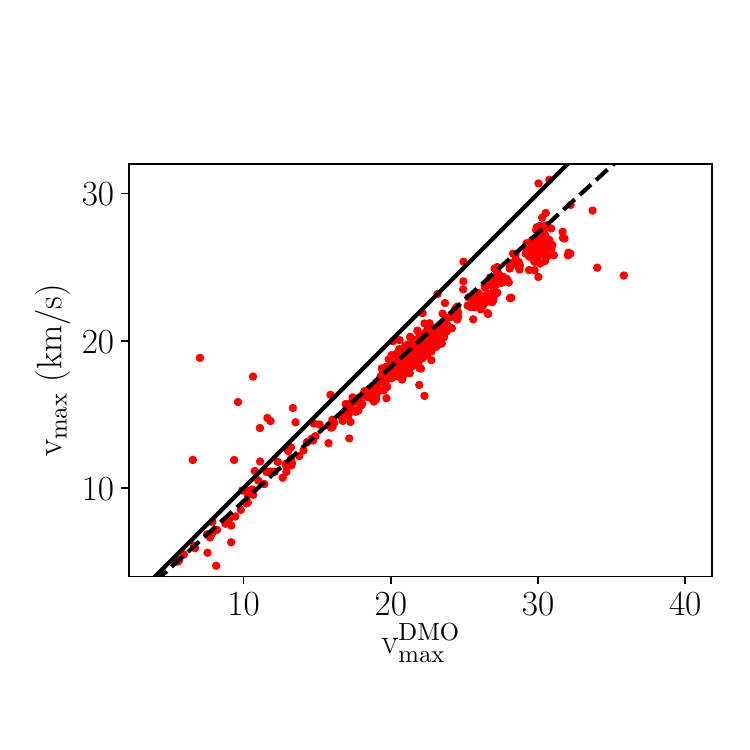}
    \caption{$v_{\rm max}$ in the dark matter only (DMO) simulations plotted against that in the baryonic simulations.  A perfect 1-to-1 match would lie along the black line; the dashed line represents multiplying the $v_{\rm max}$ from the dark matter only sims by $\sqrt{1-f_b}$, where $f_b = \Omega_b/\Omega_m$ = 0.17 is the baryon fraction of the universe. This is a better match to the $v_{\rm max}$ in the hydrodynamic simulations.
    }
    \label{fig:vmax_dmo_vs_baryons}
\end{figure}

Here, we show that a suppression in $v_{\rm max}$ from dark matter only simulations better reproduces the $v_{\rm max}$ from its corresponding hydrodynamic simulation.  In Fig. \ref{fig:vmax_dmo_vs_baryons}, we plot in red the $v_{\rm max}$ from hydrodynamic simulations ($y$-axis) against that from the DMO simulation ($x$-axis) for the same haloes.  The solid black line denotes a 1-to-1 match between the $v_{\rm max}$ in both simulations, while the dashed black line denotes a reduction in the DMO $v{\rm max}$ by a factor of $\sqrt{1-f_b}$, where $f_b$ is the baryon fraction of the universe.  We find that this reduction better matches the $v_{\rm max}$ from the hydrodynamic simulations.  The suppresion in $v_{\rm max}$ for low-mass haloes is due to loss of baryons due to reionization and stellar feedback \citep[e.g.][]{Sawala2016}.  Thus in DarkLight, we apply this correction in $v_{\rm max}$ when run on DMO simulations.

\bibliographystyle{mnras}
\bibliography{main}

\end{document}